\documentclass[12pt]{elsart}
\usepackage{latexsym}
\usepackage[final]{graphicx}
\graphicspath{{fig/}}
\newcommand{\thsp}{\thinspace}

\newcommand{\alp}{\ensuremath{\alpha}}
\newcommand{\mnuc}[2]{\ensuremath{\mathrm {^{#2}#1}}}
\newcommand{\eg}{e.g.}
\newcommand{\ttt}[1]{\ensuremath{\times 10^{#1}}}
\newcommand{\edot}{\ensuremath {\dot \epsilon}}
\newcommand{\rhoi}{\ensuremath{\rho_i}}

\newcommand{\msun}{\ensuremath {M_{\odot}}}
\newcommand{\sigv}{{\ensuremath{\langle \sigma v \rangle}}}
\newcommand{\ye}{\ensuremath{Y_{e}}}
\newcommand{\calN}{\ensuremath {\mathcal N}}
\newcommand{\calF} {\ensuremath {\mathcal {F}}}
\newcommand{\calZ} {\ensuremath {\mathcal {Z}}}
\newcommand{\calR} {\ensuremath {\mathcal {R}}}

\newcommand{\calG}{\ensuremath {\mathcal {G}}}

\newcommand{\yf}{\ensuremath {\vec {Y}^{\calF}}}
\newcommand{\yr}{\ensuremath {\vec {Y}^{\calR}}}

\newcommand{\yg}{\ensuremath {\vec {Y}^{\calG}}}
\newcommand{\ygdot}{\ensuremath {\dot {{\vec Y}^{\calG}}}}
\newcommand{\mev}{\ensuremath{\mathrm{\thsp MeV}}}
\newcommand{\erg}{\ensuremath{\mathrm{\thsp ergs}}}

\newcommand{\ergs}{\ensuremath{\mathrm{\thsp ergs/s}}}
\newcommand{\erggs}{\ensuremath{\mathrm{\thsp ergs \thsp g^{-1} \thsp s^{-1}}}}
\newcommand{\cc}{\ensuremath {\mathrm{cm^{3}}}}
\newcommand{\gcc}{\ensuremath{\mathrm{\thsp g \thsp cm^{-3}}}}

\newcommand{\gk}{\ensuremath{\mathrm{\thsp GK}}}
\newcommand{\kmps}{\ensuremath{\mathrm{\thsp km/s}}} 
\newcommand{\pergm}{\ensuremath{\mathrm{\thsp g^{-1}}}}
\newcommand{\persec}{\ensuremath{\mathrm{\thsp s^{-1}}}}     
\newcommand{\aap}{A\&A}
\newcommand{\aapr}{Astron. Astrophys. Rev.}
\newcommand{\apj}{ApJ}
\newcommand{\apjl}{ApJ Lett.}
\newcommand{\apjs}{ApJ Supp.}
\newcommand{\araa}{Ann. Rev. Ast. Ap.}
\newcommand{\arnp}{Ann. Rev. Nucl. Part. Sci.}
\newcommand{\andt}{At. Nucl. Data Tables}
\newcommand{\ausjphys}{ Aust. J. Phys.}

\newcommand{\geocosa}{Geochim. Cosmochim. Acta}
\newcommand{\jcam}{J. Comp. Appl. Math.} 
\newcommand{\jcompphys}{J. Comp. Phys.}

\newcommand{\memras}{Mem. R.A.S.}
\newcommand{\mnras}{M.N.R.A.S.}
\newcommand{\nat}{Nature}

\newcommand{\nphysa}{Nucl. Phys. A}
\newcommand{\physrep}{Phys. Rep.}

\newcommand{\prc}{Phys. Rev. C}
\newcommand{\pubastjap}{PASJ}
\newcommand{\repprogphys}{Rep. Prog. Phys.}
\newcommand{\revmodphys}{Rev. Mod. Phys.}
\newcommand{\spscirev}{Space. Sci. Rev.}
\newcommand{\zphys}{Z. Phys.}

\begin{document}
\begin{frontmatter}
%
\title{Computational Methods for Nucleosynthesis and Nuclear Energy Generation}
%
\author[AUT1,AUT2,AUT3]{W. Raphael Hix}
%
\author[AUT4,AUT3]{Friedrich-Karl Thielemann}
%
\address[AUT1]{Joint Institute for Heavy Ion Research, Oak Ridge
National Laboratory, P.O. Box 2008, Oak Ridge, TN 37831-6374}
\address[AUT2]{Department of Physics and Astronomy, University of
Tennessee, Knoxville, TN 37996-1200}
\address[AUT3]{Physics Division, Oak Ridge
National Laboratory, P.O. Box 2008, Oak Ridge, TN 37831-6373}
\address[AUT4]{Department f\"ur Physik und Astronomie, Universit\"at
Basel, CH-4056 Basel, Switzerland}
%
\begin{abstract}
This review concentrates on the two principle methods used to evolve nuclear
abundances within astrophysical simulations, evolution via rate equations and
via equilibria.  Because in general the rate equations in nucleosynthetic
applications form an extraordinarily stiff system, implicit methods have proven
mandatory, leading to the need to solve moderately sized matrix equations. 
Efforts to improve the performance of such rate equation methods are focused
on efficient solution of these matrix equations, by making best use of the
sparseness of these matrices.  Recent work to produce hybrid schemes which use
local equilibria to reduce the computational cost of the rate equations is
also discussed.  Such schemes offer significant improvements in the speed of
reaction networks and are accurate under circumstances where calculations with
complete equilibrium fail. 
\end{abstract} 
\end{frontmatter}


\section{Introduction} \label{sect:intro}

\begin{figure}[tbp]
	\centering
 	\includegraphics[angle=90,width=\textwidth]{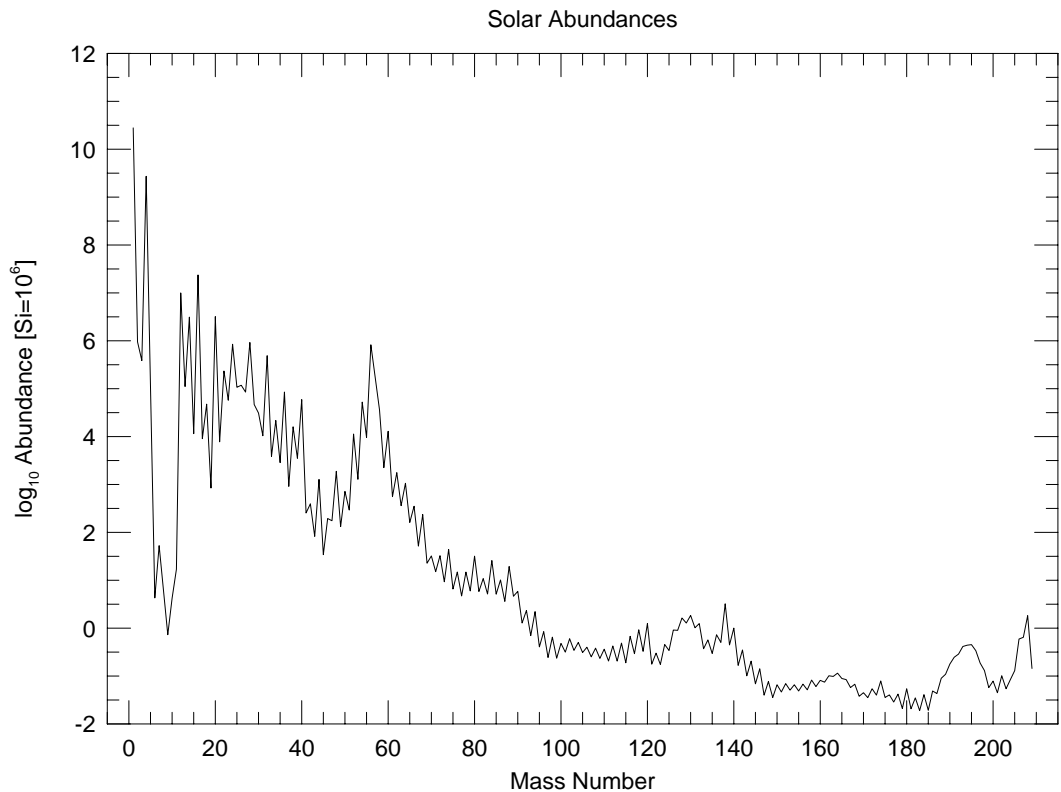}
	\caption{The abundances of isotopes in the solar system as a 
	function of atomic mass \cite{Came82,AnGr89}.  The abundances are 
	normalized so that the total abundance of silicon is $10^6$.}
	\label{fig:solab}
\end{figure}

By the second half of the last century, work by Helmholtz, Kelvin and 
others made it clear that neither gravity nor any other then known energy 
source could account for the age of the sun and solar system, as determined 
by geological measurements.  Following quickly on the heels of Rutherford's 
1911 discovery of the atomic nucleus, Eddington and others suggested that 
nuclear transmutations might be the remedy to this quandary.  With the 
burgeoning knowledge of the properties of nuclei and nuclear reactions in 
the 1930s, 1940s and 1950s, came a growing understanding of the role that 
individual nuclear reaction played in the synthesis of the elements.  In 
1957, Burbidge, Burbidge, Fowler \& Hoyle \cite{BBFH57} and Cameron 
\cite{Came57} wove these threads into a cohesive theory of nucleosynthesis, 
demonstrating how the solar isotopic abundances (displayed in 
Fig.~\ref{fig:solab}) bore the fingerprints of their astrophysical origins.  
Today, investigations refine our answers to these same two questions, how 
are the elements that make up our universe formed, and how do these 
nuclear transformations, and the energy they release, affect their 
astrophysical hosts.

In this article, we will concentrate on summarizing the two basic numerical 
methods used in nucleosynthesis studies, the tracking of nuclear 
transmutations via rate equations and via equilibria.  We will also briefly 
discuss work which seeks to meld these methods together in order to 
overcome the limitations of each.  To properly orient readers unfamiliar 
with nuclear astrophysics and to briefly describe the differing physical
conditions which influence the optimal choice of abundance evolution method, 
we begin with a brief introduction to the background astrophysics 
(\S \ref{sect:astro}), before discussing the form 
that the rate equations take (\S \ref{sect:network}).  In \S 
\ref{sect:netsolve} we will discuss the difficulties inherent in solving 
these rate equations.  \S \ref{sect:nse} describes the equations of nuclear 
equilibria as well as the limitations of their use.  Finally in \S 
\ref{sect:hyb} we will discuss hybrid schemes which seek to use local 
equilibria to simplify the rate equations.

\section{Nuclear Processes in Astrophysics} \label{sect:astro}

In general, nucleosynthesis calculations divide into two categories, (1) 
nucleosynthesis during the hydrostatic burning stages of stellar evolution  and
(2) nucleosynthesis in explosive events.  The critical distinction  between the
categories is the timescale over which transmutations occur.  In hydrostatic
burning, the release of nuclear energy occurs at a rate balancing the loss of
energy via radiation and neutrinos (the photon and neutrino luminosities),
providing continuous pressure support against the star's self-gravity.   This
results in slow burning at relatively low temperatures and densities.   In
explosive nucleosynthesis the timescale and thermodynamic conditions are 
determined by the hydrodynamics.  For example, in many cases of interest, a 
detonation shock heats the material and expels it outward, to cool 
adiabatically as it expands.  In such a case, the limiting timescale is that 
of the hydrodynamic expansion, not those intrinsic to the nuclear  processes. 
Though constrained from an in depth discussion by our  concentration on the
computational methods, in this section we will outline the  physics of the many
burning stages and in particular the way that the physics influences the choice
of method for computing the abundance changes.

\subsection{Hydrostatic Burning Stages in Stellar Evolution} \label{sect:statburn}

A star shines bright because, deep in its core, energy released by 
thermonuclear reactions balances the star's self-gravity.  
Table~\ref{tab:starlife} lists the series of burning stages which make up a 
star's life, with their representative conditions for a star like the sun 
and a star 20 times more massive (M=20 \msun).  With the consumption in 
turn of each nuclear fuel, the inexorable squeeze of gravity creates higher 
temperatures and densities.  The initiation of each subsequent stage 
requires these progressively higher temperatures and densities to overcome 
the increasing Coulomb repulsion of the reactants.  Around the core, layers 
of progressively lighter matter echo prior central burning stages.  A 
star's mass is the most important determinant of its final destiny, with 
less massive stars not progressing through the full sequence of burning 
stages.  For stars with masses less than 8 \msun, helium burning is the 
final burning stage, because the packing of electrons into a degenerate 
Fermi-Dirac configuration provides sufficient pressure support to prevent 
further contraction.  Instead, the remaining envelope is driven off, 
forming a \emph{planetary nebula}, and leaving the bare core.  Such cooling 
bare cores, composed of C \& O in this case, or He or O, Ne \& Mg in 
others, are termed \emph{white dwarf} stars.  As examination of 
Table~\ref{tab:starlife} reveals, in less massive stars the individual 
burning stages which occur do so at higher density, lower temperature and 
over much longer timescales.  For a more exhaustive description of stellar 
evolution consult, \eg, \cite{Clay83,KiWe90,Arne96}

\begin{table}[tbp]
    \centering
    \caption{Stellar Burning Stages}
    \label{tab:starlife}
    \begin{tabular}{cllllll}
      \hline
      Burning & \multicolumn{1}{c}{$\rho_c$} & \multicolumn{1}{c}{$T_c$} &
        \multicolumn{1}{c}{$\tau$}&\multicolumn{1}{c}{$L_{phot}$}&
        \multicolumn{1}{c}{$L_{\nu}$} & Primary \\
       \noalign{\vspace{-8pt}}
      Stage &\multicolumn{1}{c}{(\gcc)} &\multicolumn{1}{c}{(\gk)}& 
        \multicolumn{1}{c}{(yr)} & \multicolumn{1}{c}{(\ergs)}& 
        \multicolumn{1}{c}{(\ergs)} & Reactions \\
      \hline
      \multicolumn{7}{c}{For a 1 \msun star \cite{Baea82}} \\
      \hline
      Hydrogen & 150        & 0.015 & 1\ttt{10}  & 3.9\ttt{33} & - & 
        PP chain \\
      Helium   & 2.0\ttt{5} & 0.15  & 4 \ttt{8}  & 1.6\ttt{35} & - & 
        Triple \alp \\
      \hline
      \multicolumn{7}{c}{For a 20 \msun star \cite{ABKW89}} \\
      \hline
      Hydrogen& 5.6        & 0.040 & 1.0\ttt{7} & 2.7\ttt{38} & - & 
        CNO Cycle \\
      Helium  & 9.4\ttt{2} & 0.19  & 9.5\ttt{5} & 5.3\ttt{38} & $<\ttt{36}$ & 
        Triple \alp \\
      Carbon  & 2.7\ttt{5} & 0.81  & 3.0\ttt{2} & 4.3\ttt{38} & 7.4\ttt{39} & 
        $\mnuc{C}{12}+\mnuc{C}{12} \rightarrow \mnuc{Ne}{20} + \alpha $\\
      Neon    & 4.0\ttt{6} & 1.7   & 0.4        & 4.4\ttt{38} & 1.2\ttt{43} &
        $\mnuc{Ne}{20}+\gamma \rightarrow \mnuc{O}{16} + \alpha $\\
      Oxygen  & 6.0\ttt{6} & 2.1   & 0.5        & 4.4\ttt{38} & 7.4\ttt{43} &
        $\mnuc{O}{16}+\mnuc{O}{16}\rightarrow \mnuc{Si}{28}+\alpha$ \\
      Silicon & 4.9\ttt{7} & 3.7   & 0.01       & 4.4\ttt{38} & 3.1\ttt{45} &
        $\mnuc{Si}{28}+7 \alpha \rightarrow \mnuc{Ni}{56}$ \\
      \hline
    \end{tabular}	
\end{table}

Because the hydrostatic burning timescales are long compared to beta-decay 
half-lives (with a few exceptions of long-lived unstable nuclei), nuclei  can
decay back to stability before undergoing the next reaction.  As a result, most
reactions in hydrostatic burning stages proceed through stable nuclei.  As
indicated in Table~\ref{tab:starlife}, stars transmute hydrogen into helium 
via two alternative reaction sequences; the PP-chains, initiated by the
conversion of two protons to a positron, a  neutrino and a deuteron, (commonly
notated $\mnuc{H}{1}(p,e^+ \nu)\mnuc{H}{2}$), and the CNO cycle, which converts
$^1$H into $^4$He by a sequence of catalytic ($p,\gamma$) and ($p,\alpha$)
reactions on pre-existing C, N, and O nuclei and subsequent beta-decays.  
He-burning follows H-burning, converting \mnuc{He}{4} into \mnuc{C}{12} via 
the triple-alpha process, which we will discuss further in \S \ref{sect:hyb}. 
A portion (larger in more massive stars) of these \mnuc{C}{12} nuclei capture
an additional \alp -particle to form \mnuc{O}{16}.  With the exhaustion of
\mnuc{He}{4}, contraction continues until conditions are sufficient to fuse
pairs of \mnuc{C}{12} nuclei, producing \mnuc{Ne}{20} and a small fraction of
\mnuc{Na}{23}.  Depletion of \mnuc{C}{12} is followed by further contraction. 
Before temperatures become sufficient for \mnuc{O}{16} nuclei to fuse, the
thermal photon bath becomes energetic enough to photodisintegrate
\mnuc{Ne}{20} (see \S  \ref{sect:reacrate} and Eq.~\ref{eq:phrate} for more
details), freeing an \alp-particle which can be captured by other
\mnuc{Ne}{20} nuclei forming \mnuc{Mg}{24}.  Once \mnuc{Ne}{20} is exhausted,
continued contraction raises the temperature until it is sufficient for
\mnuc{O}{16} to fuse, producing \mnuc{Si}{28} and \mnuc{P}{31}.

Si-burning, like Ne-burning, is initiated by photodisintegration reactions 
which then provide the light particles needed for capture reactions.  
Temperatures sufficient to photodisintegrate Si are also sufficient to 
photodisintegrate all other nuclei and permit charged particle captures 
(and, in any case, neutron captures, where there is no Coulomb barrier to 
overcome).  This leaves many individual reactions in a \emph{chemical 
equilibrium} where reactions are balanced by their inverses.  If each of the 
important reactions connecting two species is in chemical equilibrium, then 
the relative abundances of these species will obey an equilibrium 
distribution.  As the most bound nuclei, and therefore most resistant to 
photodisintegration, isotopes of Fe and Ni dominate the products of silicon 
burning, producing the \emph{Fe-peak} seen near $A= 60$ in 
Fig~\ref{fig:solab}.  We will discuss this equilibrium further in \S 
\ref{sect:nse}.  For a more complete description of silicon burning, see 
\cite{WoAC73,HiTh96}.  It is important to note that because of the large 
photodisintegration fluxes which nearly balance their reverse capture 
reactions, the effective rate at which silicon is burned is as much as 
$10^{5}$ times slower than the individual reaction rates would indicate.  
This near balance of reaction rates during silicon burning will also prove 
important in our discussion of numerical methods in \S \ref{sect:hyb}.  
More extensive overviews of the major and minor reaction sequences in all 
burning stages from helium to silicon burning in massive stars is given in 
\cite{Arne96,ArTh85,ThAr85,WoWe95,Noea97}.

An additional hydrostatic nucleosynthetic process is the {\em s(low neutron 
capture)-process}.  While energetically unimportant, this slow neutron 
capture process leads to the build-up of the (small) abundances of roughly 
half of the heavy elements.  During core and shell He-burning, $(\alpha,n)$ 
reactions on neutron rich nuclei provide a source of free neutrons, 
initiating a series of neutron captures and $\beta$-decays, starting on 
pre-existing medium and heavy nuclei (up to Fe), synthesizing nuclei up to 
Pb and Bi.  Such reactions, occurring under temperature conditions where 
photodisintegration are unimportant, approach a \emph{steady flow}, wherein 
the flux of $\mnuc{Z}{A-1} (n,\gamma) \mnuc{Z}{A}$ equals the flux of 
$\mnuc{Z}{A} (n,\gamma) \mnuc{Z}{A+1}$, as long as the relevant reaction 
timescale is short in comparison to the process timescale.  For general 
overviews of the s-process see \cite{KaBW89,Kaea94,WVKK97,GaBL97}.  From a 
numerical perspective, in this and several other cases, it is often 
advantageous to couple only the dominant energy producing reactions to the 
hydrodynamics and perform detailed nucleosynthesis, which can require a 
very large nuclear network, in a post-processing fashion.

\subsection{Explosive Burning} \label{sect:expburn}

While hydrostatic sources are capable of producing many of the isotopes shown
in Fig.~\ref{fig:solab}, most of these products are trapped deep in the
potential well of their parent star or white dwarf.  Explosions are necessary
to liberate the transmuted nuclei from this gravitational embrace. In massive
stars, the formation of the iron core during silicon burning marks the end of
nuclear energy generation in the core, as nuclei more massive than the iron
peak nuclei are less bound.  Ultimately, gravitational contraction turns into
collapse, which is halted suddenly by degenerate nucleon pressure when the 
matter density approaches or exceeds that of the atomic nucleus $(\sim 10^{14}
\gcc)$.  Infalling material from overlying layers bounce off of this newly
formed \emph{proto-neutron star}, sending a shockwave outward. Though this
shock soon stalls, it is re-invigorated by neutrinos carrying off the
gravitational energy released in the formation of the proto-neutron star (see,
\eg, \cite[this volume]{Mezz99}).  This re-energized shock propagates out of the
star, unbinding much of the overlying layers and driving them into space with
velocities of thousands of \kmps, and  producing a (core-collapse) supernova
\cite{ABKW89,Woos88,Mccr93}. This passing shock also causes further
nucleosynthesis to occur as it passes through the layered composition of the 
star, raising temperatures to several \gk.  Many of the hydrostatic burning
processes discussed in \S  \ref{sect:statburn} occur under these explosive
conditions but at much  higher temperatures and over much shorter timescales. 
With little regard to the initial composition, the burning stage which
determines the nucleosynthetic outcome is that whose timescale at the peak
temperature is comparable to the hydrodynamic timescale (see, \eg,
\cite{Trur85,Trim91,Arne95}). Much more detail on the products of explosive
burning in core collapse supernova is available in the literature (see, \eg,
\cite{Arne96,WoWe95,Noea97,Arne95,ThHN90,ThNH96}).  Because these explosive 
burning stages are responsible for producing many nuclei with masses between
16-70, detailed modeling is important.  However such modeling is complicated 
by the strongly varying hydrodynamic conditions.  The computational feasibility
of performing detailed nucleosynthesis calculations (particularly explosive
silicon burning) within the latest generation of multi-dimensional 
hydrodynamic models is greatly aided by techniques like those we will discuss 
in \S \ref{sect:qse}

The explosive analogue of the s-process is the \emph{r(apid neutron
capture)-process} which requires larger neutron concentrations.  Conditions
suitable for r-process nucleosynthesis can occur in the decompression of
neutron star matter \cite{Came89,Meye89,LMRS77,ELPS89,RLTD98} or in
the innermost regions of  the core-collapse supernova ejecta
\cite{WoHo92,Hoea97}.  As a result of the electron and neutrino captures this
material has experienced, there are more neutrons than protons. With equal
numbers of protons and  neutrons tied up in \mnuc{He}{4}, the remainder provide
the required  neutron to heavy seed nucleus ratio.  Calculations independent of
the specific astrophysical site \cite{KBTM93,FKRT97}, performed with the goal
of reproducing the solar abundance pattern of heavy elements, show that
extremely unstable nuclei close to the neutron drip line are produced and
beta-decay timescales can be short in comparison to the process timescales. 
Because of the large number of nuclei involved, r-process calculations are
among the most numerically expensive, however as we will discuss in \S
\ref{sect:netsolve} \& \S \ref{sect:hyb}, partial equilibria and/or the slow
variation of the light particle abundances can be employed to simplify the
calculation.

A second type of supernova is the thermonuclear supernova, which occurs when 
explosive carbon burning is ignited in the center of an accreting white dwarf,
either as a result of the white dwarf exceeding the maximum mass ($\sim 1.4
\msun$) which can be supported by electron degeneracy pressure (see, \eg, 
\cite{Chan35,NoTY84,NiWo97,KhOW97,HoWT98}) or due to compression resulting from the
detonation of an accreted He layer (see, \eg, \cite{LiAr95,Woos97a}). In either
case, a flame front propagates outward disrupting the white dwarf and leaving 
a composition dominated by iron peak and intermediate mass nuclei. 
Computationally, in addition to sharing the complications discussed for
explosive burning in core collapse supernovae, explosive nucleosynthesis in
these thermonuclear supernovae is also the source of the energy which powers
the explosion. Thus accurate hydrodynamic simulation requires at least the
inclusion of an accurate means to calculate the rate of thermonuclear energy
release. 

If a white dwarf accretes hydrogen (typically via mass transfer from a  binary
companion) slowly enough, a layer of unburned hydrogen will build up  on the
surface of the white dwarf.  Once the density of this layer is  sufficient to
ignite the hydrogen, a \emph{nova} results (see, \eg, 
\cite{SuNo80,Stea93,Coea95}).  The degeneracy of the material and the steep 
gravitational gradients at the surface of the white dwarf, result in  explosive
hydrogen burning via the ``hot'' or \emph{$\beta$-limited} CNO cycle
\cite{RFRT97,JoHe98},  releasing $10^{46}-10^{47} \erg$ over timescales of
100-1000 s, with peak  temperatures reaching $0.2-0.3 \gk$.  A  neutron star
can also accrete mass from a companion in a similar fashion, building up a
layer of hydrogen which explodes to produce an \emph{X-ray  burst} (see, \eg,
\cite{WaWo81,TaWL96}).  Because the neutron stars surface  gravitational
gradients are even stronger, the timescales are shorter (1-10  s) and even
higher peak temperatures ($1-2 \gk$) are reached, enabling  proton capture up
to and beyond the Fe-peak, the \emph{r(apid) p(roton capture)-process}
\cite{RFRT97,SAGW98}. However the size of the hydrogen layer is much smaller
(as small as $10^{-12} \msun$) and, as a result, so is the energy release,
which is typically  $10^{39}-10^{40} \erg$.  The thermonuclear source of the
explosive energy  in novae and X-ray burst, as well as the similarity in the
convective and  nuclear timescales, require that hydrodynamic simulations of
these objects  include large networks.  Fortunately, approximations (which we
will discuss  in \S \ref{sect:hyb}) exist which can greatly reduce the
computational  cost.

An additional form of explosive burning occurred as the universe expanded 
from its primordial \emph{Big Bang}.  Since its origin, the universe has 
been expanding, cooling from the initially extreme temperatures.  At the 
earliest times, the populations of all kinds of sub-atomic particles were 
equilibrated.  Eventually, continued cooling and the freezeout of this 
equilibrium left only a few neutrons and protons (approximately one baryon 
per billion photons) with approximately 7 protons per neutron.  These 
nucleons largely remained free because the small Q-value of the reaction $n 
+p \rightleftharpoons \mnuc{H}{2} + \gamma$ permits chemical equilibrium to 
persist to temperatures of 1 \gk, keeping the abundance of deuterium very 
small.  By the time the universe has cooled to 1 \gk, the expansion has 
reduced the density to $\sim 10^{-5} \gcc$, a density small enough that, 
even with an expansion timescale of days, the resulting tiny flux through 
the $3 \alpha$ reaction did not produce significant amounts of heavy 
elements.  Instead Big Bang nucleosynthesis is responsible for the 
production of the lightest elements; \mnuc{H}{1}, \mnuc{He}{4}, \mnuc{H}{2}, 
\mnuc{He}{3} and \mnuc{Li}{7} \cite{Wago73,ThSO94}.  In principle, the 
small number of affected species and the simple hydrodynamic evolution make 
Big Bang nucleosynthesis the least computationally challenging.  However 
the greater accuracy of the relevant nuclear reactions, as well as the 
important limits that Big Bang nucleosynthesis calculations place on 
cosmologically important factors, have resulted in a much stronger drive to 
high precision in Big Bang nucleosynthesis calculations \cite{SmKM93,OlSS97}.

\section{Thermonuclear Reaction Networks} \label{sect:network}

From the discussion in the preceding section, it is clear that nuclear 
abundances in many cases obey equilibrium distributions.  But the general 
case requires the evolution of nuclear abundances via a nuclear reaction 
network.  Composed of a system of first order differential equations, the 
nuclear reaction network has sink and source terms representing each of the 
many nuclear reactions involved.  Prior to discussing the numerical 
difficulties posed by the nuclear network, it is necessary to understand 
the sets of equations we are attempting to solve.  To this end, we present 
a brief overview of the thermonuclear reaction rates of interest and how 
these rates are assembled into the differential equations we must 
ultimately solve.  For more detailed information, we refer the reader to a 
number of more complete discussions 
\cite{Clay83,Arne96,RoRo88,Woos86,ThNH94} which can be found in the 
literature.  We will end this section by briefly discussing the coupling 
of nucleosynthesis with hydrodynamics.

\subsection{Thermonuclear Reaction Rates} \label{sect:reacrate}

There are a large number of types of nuclear reactions which are of 
astrophysical interest.  In addition to the emission or absorption of  nuclei
and nucleons, nuclear reactions can involve the emission or  absorption of
photons ($\gamma$ -rays) and leptons (electrons, neutrinos,  and their
anti-particles).  As a result, nuclear reactions involve three of the  four
fundamental forces, the nuclear strong, electromagnetic and nuclear weak 
forces.  Reactions involving leptons (termed weak interactions) proceed  much
more slowly than those involving only nucleons and photons.  However  these
reactions are still important, as only weak interactions can change the global
ratio of protons to neutrons.

The most basic piece of information about any nuclear reaction is the
nuclear cross section.  The cross section for a reaction between target 
$j$ and projectile $k$ is defined by 
\begin{equation}
    \sigma = {\rm{number\ of\ reactions\ target^{-1} sec^{-1}} \over
    {flux\ of\ incoming\ projectiles}} = {{r/n_j} \over {n_k v}}. 
    \label{eq:sigma}
\end{equation}
The second equality holds when the relative velocity between targets of 
number density $n_j$ and projectiles of number density $n_k$ is constant 
and has the value $v$.  Then $r$, the number of reactions per \cc\ and sec, 
can be expressed as $r=\sigma v n_j n_k$.  More generally, the targets and 
projectiles have distributions of velocities, in which case $r$ is given by
\begin{equation}
    r_{j,k}=\int \sigma (\vert \vec v_j -\vec v_k\vert) \vert \vec v_j 
    -\vec v_k\vert d^3 n_j d^3 n_k.
    \label{eq:rate}
\end{equation}
The evaluation of this integral depends on the types of particles and
distributions which are involved. For nuclei $j$ and $k$ in an astrophysical 
plasma, Maxwell-Boltzmann statistics generally apply, thus
\begin{equation}
    d^3n=n ({{m} \over {2\pi k_B T}})^{3/2} \exp (-{{mv^2} \over {2 k_B T}}) d^3v,
\end{equation}
allowing $n_j$ and $n_k$ to be moved outside of the integral.  
Eq.~\ref{eq:rate} can then be written as $r_{j,k}=\sigv_{j,k} n_j n_k$, 
where \sigv\ is the velocity integrated cross section.  Equivalently, one 
can express the reaction rate in terms of a mean lifetime of particle $j$
against destruction by particle $k$, 
\begin{equation}
    \tau_{k}(j)= {1 \over {\sigv_{j,k} n_{k}}}
    \label{eq:tau}
\end{equation}
For thermonuclear reactions, these integrated cross sections have the form 
\cite{Clay83,FoCZ67}
\begin{equation}
    \langle j,k \rangle \equiv \sigv_{j,k}= ({8 \over {\mu \pi}})^{1/2} 
    (k_B T)^{-3/2} \int_0 ^\infty E \sigma (E) {\rm exp}(-E/k_B T) dE, 
    \label{eq:sigv}
\end{equation}
where $\mu$ denotes the reduced mass of the target-projectile system, $E$ 
the center of mass energy, $T$ the temperature and $k_{B}$ is Boltzmann's 
constant.

Experimental measurements and theoretical predictions for these reaction  rates
provide the data input necessary for nuclear networks.  While detailed 
discussion of individual rates is beyond the scope of this article, the 
interested reader is directed to the following reviews.  Experimental  nuclear
rates have been reviewed in detail by  \cite{RoRo88,Waea97,KaTW98}. The most
recent experimental  charged particle rate compilations are the ones by
\cite{CaFo88,ArGJ99}. Experimental neutron capture cross sections are
summarized by  \cite{BaKa87,BeVW92,WVKK97}. Rates for unstable (light) nuclei 
are given, for example, by \cite{SAGW98,ThSO94,WiGT88,WGGB89,TSOF93,Raea94}. 
For the vast number of  medium and heavy nuclei which exhibit a high density of
excited states at  capture energies, Hauser-Feshbach (statistical model)
calculations are  applicable. The most recent compilations were provided by 
\cite{HWFZ76,WFHZ78,ThAT87,CoTT91}.   Improvements in level densities
\cite{RaTK97}, alpha potentials, and the consistent treatment of isospin mixing
will lead to the next generation of theoretical rate predictions
\cite{RaTh98,SFKR98}.

In practice, these experimental and theoretical reaction rates are determined 
for bare nuclei, while in astrophysical plasmas, these reactions occur 
among a background of other nuclei and electrons.  As a result of this 
background, the reacting nuclei experience a Coulomb repulsion modified from 
that of bare nuclei. For high densities and/or low temperatures, the effects 
of this screening of reactions becomes very important. Under most 
conditions (with non-vanishing temperatures) the generalized reaction rate 
integral can be separated into the traditional expression without screening 
[Eq.~\ref{eq:sigv}] and a screening factor, 
\begin{equation}
    \langle j,k \rangle^*=f_{scr}(Z_j,Z_k,\rho,T,n_i) \langle j,k \rangle. 
    \label{eq:screen}
\end{equation}
This screening factor is dependent on the charge of the involved particles,
the density, temperature, and the composition of the plasma. For more details 
on the form of $f_{scr}$, see, \eg, \cite{Salp54,SavH69,ThTr87,Ichi93,BrSa97}.  
At high densities and low temperatures screening factors can enhance 
reactions by many orders of magnitude and lead to {\em pycnonuclear ignition}.
In the extreme case of very low temperatures, where reactions are only
possible via ground state oscillations of the nuclei in a Coulomb lattice,
Eq.~\ref{eq:screen} breaks down, because it was derived under the assumption 
of a Boltzmann distribution (for recent references, see \cite{BrSa97,Ichi96}).

When particle $k$ in Eq.~\ref{eq:rate} is a photon, the distribution 
$d^3n_k$ is given by the Plank distribution, 
\begin{equation}
    d^3n_{\gamma} =  {{8 \pi} \over {c^3 h^3}} {{E_{\gamma}^2} \over
    {\exp \left(E_{\gamma} /k_B T\right)-1}} dE_{\gamma} \ .
\end{equation}
Furthermore, the relative velocity is always $c$ and thus the integral 
is separable, simplifying to 
\begin{equation}
    r_j  = {{\int d^3n_j}\over {\pi ^2 (c \hbar )^3}} \int_0 ^\infty {{c 
    \sigma(E_\gamma ) E_{\gamma}^2} \over {{\rm exp}(E_\gamma /k_B T) -1}} dE_
    \gamma \equiv \lambda_{j,\gamma} (T) n_j. 
    \label{eq:phrate}
\end{equation}
In practice there is, however, no need to directly evaluate the 
photodisintegration cross sections, because they can be expressed by detailed
balance in terms of the capture cross sections for the inverse 
reaction, $l+m\rightarrow j+\gamma$ \cite{FoCZ67}.
\begin{equation}
    \lambda_{j,\gamma} (T)= ({{G_l G_m} \over G_j}) ({{A_l A_m} \over A_j})
    ^{3/2} ({{m_u k_B T} \over {2\pi \hbar^2}})^{3/2} \langle l,m \rangle 
    \exp (-Q_{lm}/k_B T).
    \label{eq:detbal}
\end{equation}
This expression depends on the partition functions, $G_{k}=\sum_i (2J_i+1) 
\exp (-E_i/k_B T)$ (which account for the populations of the excited 
states of the nucleus), the mass numbers, $A$, the temperature 
$T$, the inverse reaction rate $\langle l,m \rangle$, and the reaction 
$Q$-value (the energy released by the reaction), $Q_{lm}=(m_l+m_m-m_j) 
c^2$.  Since photodisintegrations are endoergic, their rates are 
vanishingly small until sufficient photons exist in the high energy tail of 
the Planck distribution with energies $>Q_{{lm}}$.  As a rule of thumb this 
requires $T$$\approx$$Q_{lm}/30 k_{B}$.

A procedure similar to that for Eq.~\ref{eq:phrate} applies to captures of 
electrons by nuclei.  Because the electron is 1836 times less massive than a 
nucleon, the velocity of the nucleus $j$ in the center of mass system is 
negligible in comparison to the electron velocity ($\vert \vec v_j- \vec 
v_e \vert \approx \vert \vec v_e \vert$).  In the neutral, completely 
ionized plasmas typical of the astrophysical sites of nucleosynthesis, the 
electron number density, $n_{e}$, is equal to the total density of protons 
in nuclei, $\sum_i Z_i n_i$.  However in many of these astrophysical 
settings the electrons are at least partially degenerate, therefore the 
electron distribution cannot be assumed to be Maxwellian.  Instead the 
capture cross section has to be integrated over a Boltzmann, partially 
degenerate, or degenerate Fermi distribution of electrons, depending on the 
astrophysical conditions.  The resulting electron capture rates are 
functions of $T$ and $n_e$,
\begin{equation}
    r_j=\lambda_{j,e} (T,n_e) n_j. 
    \label{eq:ecrate}
\end{equation}
Similar equations apply for the capture of positrons which are in thermal 
equilibrium with photons, electrons, and nuclei.  Electron and positron 
capture calculations have been performed by \cite{FuFN80,FuFN82,FuFN85} for a
large variety of nuclei with mass numbers between A=20 and A=60.  For 
improvements and application to heavier nuclei see also
\cite{Taea89,Deea98,AFWH94,SuSR97}. 

For normal decays, like beta or alpha decays, with a characteristic  half-life
$\tau_{1/2}$,  Eq.~\ref{eq:phrate}~\&~\ref{eq:ecrate} also applies, with the
decay constant  $\lambda_j=\ln 2/\tau_{1/2}$.  In addition to innumerable
experimental  half-life determinations, beta-decay half-lives for unstable
nuclei have  been predicted by \cite{TaYK73,KlMO84} \& \cite[including
temperature  effects]{TaYo88}.  More recently, estimates have been made with
improved  quasi particle RPA calculations 
\cite{SBMK89,SBMK90,MoRa90,HSMK92,PfKr96,MoNK97,Borz97,OdHM94}.

At high densities ($\rho \sim 10^{13} \gcc$), even though the size of 
the neutrino scattering cross section on nuclei and electrons is very 
small, enough scattering events occur to thermalize the neutrino 
distribution.  Under such conditions the inverse process to electron 
capture (neutrino capture) can occur in significant numbers and the 
neutrino capture rate can be expressed in a form similar to 
Eqs.~\ref{eq:phrate}~\&~\ref{eq:ecrate} by integrating over the thermal 
neutrino distribution (e.g. \cite{FuMe95}).  Inelastic neutrino scattering 
on nuclei can also be expressed in this form.  The latter can cause 
particle emission, similar to photodisintegration (e.g. 
\cite{WHHH90,KKLT92,KLKT93,KLTV95,Qiea96}).  The calculation of these rates 
can be further complicated by the neutrinos not being in thermal 
equilibrium with the local environment.  When thermal equilibrium among 
neutrinos was established at a different location, then the neutrino 
distribution might be characterized by a chemical potential and a 
temperature different from the local values.  Otherwise, the neutrino 
distribution must be evolved in detail \cite[this volume]{Mezz99}.

\subsection{Thermonuclear Rate Equations} \label{sect:yderiv}

The large number of reaction types discussed in \S \ref{sect:reacrate} can 
be divided into 3 functional categories based on the number of reactants 
which are nuclei.  The reactions involving a single nucleus, which include 
decays, electron and positron captures, photodisintegrations, and neutrino 
induced reactions, depend on the number density of only the target species.  
For reaction involving two nuclei, the reaction rate depends on the number 
densities of both target and projectile nuclei.  There are also a few 
important three-particle process, like the triple-\alp\ process discussed 
in \S \ref{sect:hyb}, which are commonly successive captures with an 
intermediate unstable target (see, \eg, \cite{NoTM85,GoWT95}).  Using an 
equilibrium abundance for the unstable intermediate, the contributions of 
these reactions are commonly written in the form of a three-particle 
processes, depending on a trio of number densities.  Grouping reactions by 
these 3 functional categories, the time derivatives of the number densities 
of each nuclear species in an astrophysical plasma can be written in terms 
of the reaction rates, $r$, as
\begin{equation}
    \left. {{\partial n_i} \over {\partial t}} \right|_{\rho =const}= \sum_j 
    \calN^i _j r_j + \sum_{j,k} \calN^i _{j,k} r_{j,k} + \sum_{j,k,l} 
    \calN^i _{j,k,l} r_{j,k,l},
    \label{eq:ndot}
\end{equation}
where the three sums are over reactions which produce or destroy a nucleus 
of species $i$ with 1, 2 \& 3 reactant nuclei, respectively.  The \calN\ s
provide for proper accounting of numbers of nuclei and are given by: 
$\calN^i_j = N_i$, $\calN^i_{j,k} = N_i / \prod_{m=1}^{n_m} | N_{j_m} |!  
$, and $\calN^i_{j,k,l} = N_i / \prod_{m=1}^{n_m} |N_{j_m}|!$.  The $N_i's$ 
can be positive or negative numbers that specify how many particles of 
species $i$ are created or destroyed in a reaction, while the denominators, 
including factorials, run over the $n_m$ different species destroyed in the 
reaction and avoid double counting of the number of reactions when 
identical particles react with each other (for example in the \mnuc{C}{12} + 
\mnuc{C}{12} or the triple-\alp\ reactions; for details see \cite{FoCZ67}).

In addition to nuclear reactions, expansion or contraction of the plasma 
can also produce changes in the number densities $n_i$.  To separate the 
nuclear changes in composition from these hydrodynamic effects, we 
introduce the nuclear abundance $Y_i =n_i/\rho N_A$, where $N_{A}$ is 
Avagadro's number.  For a nucleus with atomic weight $A_i$, $A_iY_i$ 
represents the mass fraction of this nucleus, therefore $\sum A_iY_i=1$.  
Likewise, the equation of charge conservation becomes $\sum Z_i Y_i = Y_e$, 
where $\ye (= n_e /\rho N_A)$ is the electron abundance.  By recasting 
Eq.~\ref{eq:ndot} in terms of nuclear abundances $Y_i$, a set of ordinary 
differential equations 
for the evolution of $\dot Y_i$ results which depends only on nuclear 
reactions.  In terms of the reaction cross sections introduced in \S 
\ref{sect:reacrate}, this reaction network is described by the following 
set of differential equations
\begin{eqnarray}
    \dot Y_i & = & \sum_j \calN^i _j \lambda_j Y_j + \sum_{j,k} \calN^i _{j,k} 
    \rho N_A \langle j,k \rangle Y_j Y_k  \nonumber \\ 
    & + & \sum_{j,k,l} \calN^i _{j,k,l} \rho^2 N_A^2 \langle j,k,l \rangle 
	Y_j Y_k Y_l. 
    \label{eq:ydot}
\end{eqnarray}

\subsection{Coupling Nuclear Networks to Hydrodynamics}

As we touched on in the previous section, nuclear processes are tightly linked
to the hydrodynamic behavior of the bulk medium.  Thermonuclear  processes
release (or absorb) energy, altering the pressure and  causing hydrodynamic
motions.  These motions may disperse the thermonuclear  ash, bringing a
continued supply of fuel to support the flame.  The  compositional changes,
both of nuclei and of leptons, caused by  thermonuclear reactions can also
change the equation of state and opacity,  further impacting the hydrodynamic
behavior.  For purposes of this review  of nucleosynthesis methods, which
generally assume that thermonuclear and  hydrodynamic changes in local
composition can be successfully decoupled, we  include a brief description of
how this decoupling is best achieved.   M\"uller \cite{Muel98} provides an
authoritative overview and discusses the difficulties (and open issues)
involved when including nucleosynthesis within hydrodynamic  simulations.

The coupling between thermonuclear processes and hydrodynamic changes can 
be divided into two categories by considering the spatial extent of the 
coupling.  Nucleosynthetic changes in composition and the resultant energy 
release produce \emph{local} changes in hydrodynamic quantities like 
pressure and temperature.  The strongest of these local couplings is the 
release (or absorption) of energy and the resultant change in temperature.  
Changes in temperature are particularly important because of the 
exponential nature of the temperature dependence of thermonuclear reaction 
rates.  Since the nuclear energy release is uniquely determined by the 
abundance changes, the rate of thermonuclear energy release, $\dot 
\epsilon$, is given by
\begin{equation}
    \dot \epsilon_{nuc} = - \sum_i  N_A M_i c^2 \dot Y_i (\mev \pergm \persec).
    \label{eq:edot}
\end{equation}
where $M_i c^2$ is the rest mass energy of species $i$ in \mev.  Since all 
reactions conserve nucleon number, the atomic mass excess $M_{ex,i}=M_i - 
A_i m_u$ ($m_u$ is the atomic mass unit) can be used in place of the mass 
$M_i$ in Eq.~\ref{eq:edot} (see \cite{AuWa95} for a recent compilation of 
mass excesses).  The use of atomic mass units has the added benefit that 
electron conservation is correctly accounted for in the case of $\beta^{-}$ 
decays and $e^{-}$ captures, though reactions involving positrons require 
special treatment.  In general, the nuclear energy release is deposited 
locally, so the rate of thermonuclear energy release is equal to the 
nuclear portion of the hydrodynamic heating rate.  However, there are 
instances where nuclear products do not deposit their energy locally.  
Escaping neutrinos can carry away a portion of the thermonuclear energy 
release.  In the rarefied environment of supernova ejecta at late times, 
positrons and gamma rays released by $\beta$ decays are not completely 
trapped. In most such cases, the escaping particles stream freely 
from the reaction site, allowing adoption of a simple loss term analogous to 
Eq.~\ref{eq:edot} with $M_i c^2$ replaced by an averaged energy loss term.  
For example, the weak reaction rate tabulations of Fuller, Fowler, \& 
Neumann \cite{FuFN85} provide averaged neutrino losses.  From these we can 
construct
\begin{equation}
    \dot \epsilon_{\nu \ loss} =  \sum_i \langle E_{\nu} \rangle \dot 
    Y_{i,weak} ,
    \label{eq:enudot}
\end{equation}
where we consider only those contributions to $\dot Y$ due to neutrino 
producing reactions.  In some cases, like supernova core collapse (see 
\cite{Mezz99}, this volume), more complete transport of the escaping 
leptons or gamma rays must be considered.  Other important quantities which 
are impacted by nucleosynthesis, like \ye, can be obtained by appropriate 
sums over the abundances and also need not be evolved separately.

Implicit solution methods require the calculation of $\dot Y (t+\Delta t)$, 
where $\Delta t$ is the nuclear timestep, which in turn requires knowledge  of
$T(t+\Delta t)$.  One could write a differential equation for the energy 
release analogous to Eq.~\ref{eq:ydot}, with the \calN s replaced by the 
reaction $Q$-values, and thereby evolve the energy release (and calculate 
temperature changes) as an additional equation within the network solution.  
M\"uller \cite{Muel86} has shown that such a scheme can help avoid 
instabilities in the case of a physically isolated zone entering or leaving 
nuclear statistical equilibrium.  In general, however, use of this additional
equation is made unnecessary by the relative slowness with which the
temperature changes.  The timescale on which the temperature changes is given by
\begin{equation}
    \tau_{T}= T / {\dot T} \approx C_{V}T/ \dot \epsilon_{nuc}
    \label{eq:taut}
\end{equation}
and is often called the \emph{ignition timescale}.  The timescale on which an 
individual abundance changes is its \emph{burning time}, 
\begin{equation}
    \tau (\mnuc{Z}{A}) = Y(\mnuc{Z}{A}) /{\dot Y(\mnuc{Z}{A})} = \min_{k} 
    {\tau_{k}(\mnuc{Z}{A})} 
    \label{eq:tauy}
\end{equation}
where $\tau_{k}(\mnuc{Z}{A})$ is defined in Eq.~\ref{eq:tau}.  In general 
$\tau_{T}$ differs from $\tau (\mnuc{Z}{A})$ of the principle fuel by the 
ratio of thermal energy content to the energy released by the reaction.  
For degenerate matter this ratio can approach zero, allowing for explosive 
burning.  In contrast, best results for the nuclear network are achieved 
\cite{ArTr69} when the network timestep $\Delta t$ is chosen to be the 
burning timescale of a less abundant species, typically with an abundance 
of $10^{-6}$ or smaller.  Since the dominant fuel is typically one of the 
more abundant constituents and the burning timescales are proportional to 
the abundance, $\tau_{T}$ is typically an order of magnitude or more larger 
than the network timestep (see, \eg, \cite{WeZW78,BeHT89}.  It is therefore
sufficient to calculate the  energy gain at the end of a timestep via
equation~\ref{eq:edot}, modified  as discussed above, and approximate
$T(t+\delta t) \approx T(t)$ or to  extrapolate based on \edot(t).  Since other
locally coupled quantities have  characteristic timescales much longer than
$\Delta t$, they too can be  decoupled in a similar fashion.  For the remainder
of this review, we will  consider only the equations governing changes in
isotopic abundances,  remembering that additional equations can easily be
constructed for those  special circumstances where they are necessary. 

Spatial coupling, particularly the modification of the composition by 
hydrodynamic movements such as diffusion, convective mixing and advection 
(in the case of Eulerian hydrodynamics methods), represents a more 
difficult challenge.  By necessity, an individual nucleosynthesis 
calculation examines the abundance changes in a locality of uniform 
composition.  The difficulties associated with strong spatial coupling of 
the composition occur because this nucleosynthetic calculation is spread 
over an entire hydrodynamic zone.  Convection can result in strong 
abundance gradients across a single hydrodynamic zone, which with the 
assumption of compositional uniformity, can result in very different 
outcomes as a function of the fineness of the hydrodynamic grid.  Eulerian 
advection of compositional boundaries can also have extremely unphysical 
consequences.  Fryxell \etal\ \cite{FrMA89} demonstrated how this artificial 
mixing can produce an unphysical detonation in a shock tube calculation by 
mixing cold unburnt fuel into the hot burnt region.  A related problem is 
the conservation of species.  Hydrodynamic schemes must carefully conserve 
the abundances (or partial densities) of all species 
\cite{FrMA89,Larr91,PlMu99}, lest they provide unphysical abundances to the 
nucleosynthesis calculations, which must assume conservation, and thereby 
produce unphysical results.  Because of these problems, nucleosynthesis 
calculations are best suited to hydrodynamic simulations with excellent 
capture of shock and contact discontinuities.

The relative size of the burning timescales, when compared to the relevant 
diffusion, sound crossing or convective timescale, dictates how these 
problems must be addressed.  If all of the burning timescales are much 
shorter than the timescale on which the hydrodynamics changes the 
composition, then the assumption of uniform composition is satisfied and 
the nucleosynthesis of each hydrodynamic zone can be treated independently.  
If all of the burning timescales are much larger than, for example, the 
convective timescale, then the composition of the entire convective zone 
can be treated as uniform and slowly evolving.  The greatest complexity 
occurs when the timescales on which the hydrodynamics and nucleosynthesis 
change the composition are similar.  Oxygen shell burning represents an 
excellent example of this as the sound travel, convective turnover and 
nuclear burning timescales are all of the same order as the evolutionary 
time.  The results (of 2D simulations \cite{BaAr98}) demonstrate convective 
overshooting, highly non-uniform burning and a velocity structure dominated 
by convective plumes.  Silicon burning also represents a particular 
challenge \cite{Arne96}, as the timescales for the transformation of 
silicon to iron are much slower than the convective turnover time, but the 
burning timescales for the free neutrons, protons and $\alp$-particles 
which maintain QSE are much faster, providing a strong motivation for the 
hybrid networks we will discuss in \S \ref{sect:hyb}.

\section{Solving the Nuclear Network} \label{sect:netsolve}

In principle, the initial value problem presented by the nuclear network 
can be solved by any of a large number of methods discussed in the 
literature.  However the physical nature of the problem, reflected in the 
$\lambda$'s and \sigv 's, greatly restricts the optimal choice.  The large 
number of reactions display a wide range of reaction timescales, $\tau$ 
(see Eq.~\ref{eq:tau}).  Systems whose solutions depend on a wide range of 
timescales are termed \emph{stiff}.  Gear \cite{Gear71} demonstrated that 
even a single equation can be stiff if it has both rapidly and slowly varying 
components.  Practically, stiffness occurs when the limitation of the 
timestep size is due to numerical stability rather than accuracy.  A more 
rigorous definition \cite{Lamb80} is that a system of equations $\vec{\dot Y}
(\vec Y)$ is stiff if the eigenvalues $\lambda_{j}$ of the Jacobian 
$\partial \vec{\dot Y} / \partial \vec Y$ obey the criteria
\begin{eqnarray}
    \Re(\lambda_{j}) &<& 0, \qquad j=1,\cdots,N  \\
    {\mathcal S} &=& {{max |\Re(\lambda_{j})|} \over {min |\Re(\lambda_{j})|}} 
    \gg 1 \nonumber
    \label{eq:lambdastiff}
\end{eqnarray}
where $\Re(\lambda)$ is the real part of the eigenvalues $\lambda$.  As we will
explain in this section, $\mathcal S > 10^{15}$ is not uncommon in
astrophysics.

Nucleosynthesis calculations belong to the more general field of
reactive flows, and therefore share some characteristics with related
terrestrial fields.  In particular, chemical kinetics, the study of the
evolution of chemical abundances, is an important part of atmospheric and
combustion physics and produces sets of equations much like Eq.~\ref{eq:ndot}
(see \cite{OrBo87} for a good  introduction).  These chemical kinetics systems
are known for their stiffness and a great deal of effort has been expended on
developing methods to solve these equations.  Many of the considerations for
the choice of solution method for chemical kinetics also 
apply to nucleosynthesis calculations.  In both cases, temporal integration of
the reaction rate equations is broken up into short intervals because of the
need to update the hydrodynamics variables.  This favors one step, self
starting algorithms.  Because abundances must be tracked for a large number of
computational cells (hundreds to thousands for one dimensional models,
millions for the coming generation of three dimensional models), memory storage
concerns favor low order methods since they don't require  the storage of as
much data from prior steps.  In any event, both the  errors in fluid dynamics
and in the reaction rates are typically a few  percent or more, so the greater
precision of these higher order methods often  does not result in greater
accuracy.

Because of the wide range in timescales between strong, electromagnetic and 
weak reactions, the nuclear networks are extraordinarily stiff.  PP chain
nucleosynthesis, responsible for the energy output of the Sun, offers an
excellent example of the difficulties.  The first reaction of the PP1 chain is
$\mnuc{H}{1}(p,e^{+} \nu)  \mnuc{H}{2}$, the fusion of two protons to form
deuterium.  This is a weak  reaction, requiring the conversion of a proton into
a neutron, and  releasing a positron and a neutrino.  As a result, the reaction
timescale  $\tau_{p}(\mnuc{H}{1})$ is very long, billions of years for
conditions like  those in the solar interior.  The second reaction of the PP1
chain is the  capture of a proton on the newly formed deuteron,
$\mnuc{H}{2}(p,\gamma)  \mnuc{He}{3}$.  For conditions like those in the solar
interior, the  characteristic timescale, $\tau_{p}(\mnuc{H}{2})$ is a few
seconds.  Thus  the timescales for two of the most important reactions for
hydrogen burning  in stars like our Sun differ by more than 17 orders of
magnitude (see  \cite{Clay83} for a more complete discussion of the PP chain). 
This  disparity results not from a lack of p+p collisions (which occur at a
rate  $Y(\mnuc{H}{1})/ Y(\mnuc{H}{2}) \sim 10^{17}$ times more often than 
\mnuc{H}{1} + \mnuc{H}{2} collisions), but from the rarity of the (weak)
transformation of a proton to a neutron.  While the presence of weak  reactions
among the dominant energy producing reactions is unique to hydrogen burning,
most nucleosynthesis calculations are similarly stiff, in part because of the
need to include weak interactions but also the potential for neutron capture
reactions, which occur very rapidly even at low temperature, following any
release of free neutrons.  Though further  investigation is warranted, the
nature of the nuclear reaction network  equations has thus far limited the
astrophysical usefulness of the most  sophisticated methods to solve stiff
equations developed for chemical  kinetics.

For a set of nuclear abundances $\vec Y$, one can calculate the time 
derivatives of the abundances, $\dot {\vec Y}$ using Eq.~\ref{eq:ydot}.   The
desired solution is the abundance at a future time, $\vec Y(t+\Delta  t)$,
where $\Delta t$ is the network timestep.  Since coupling with hydrodynamics 
favors low order, one step methods, general nucleosynthesis calculations  use
the simple finite difference prescription
\begin{equation}
	{{\vec Y(t+\Delta t)- \vec Y(t)} \over {\Delta t}} = (1-\Theta) \dot 
	{\vec Y}(t+\Delta t) + \Theta \dot {\vec Y}(t).
    \label{eq:deriv}
\end{equation}
With $\Theta=1$, Eq.~\ref{eq:deriv} becomes the explicit Euler method 
while for $\Theta=0$ it is the implicit backward Euler method, both of 
which are first order accurate.  For $\Theta=1/2$, Eq.~\ref{eq:deriv} is 
the semi-implicit trapezoidal method, which is second order accurate.  For 
the stiff set of non-linear differential equations which form most nuclear 
networks, a fully implicit treatment is generally most successful 
\cite{ArTr69}, though the semi-implicit method has been used in Big Bang 
nucleosynthesis calculations \cite{Wago73}, where coupling to hydrodynamics 
is less important.  Solving the fully implicit version of Eq.~\ref{eq:deriv} 
is equivalent to finding the zeros of the set of equations
\begin{equation}
    \vec \calZ(t+\Delta t)\equiv {{\vec Y(t+\Delta t)- \vec Y(t)} \over 
    {\Delta t}} - \dot {\vec Y}(t+\Delta t) =0 \ .
	\label{eq:zer}
\end{equation}
This is done using the Newton-Raphson method (see, \eg, \cite{NumRec}), 
which is based on the Taylor series expansion of $\vec \calZ(t+\Delta t)$, 
with the trial change in abundances given by
\begin{equation}
	\Delta \vec Y = \left( \partial \vec \calZ (t+\Delta t) \over 
	\partial \vec Y (t+\Delta t) \right)^{-1} \vec \calZ \ ,
	\label{eq:dely}
\end{equation}
where $\partial \vec \calZ / \partial \vec Y $ is the Jacobian of $\vec 
\calZ$.  Iteration continues until $\vec Y(t+\Delta t)$ converges.

A potential numerical problem with the solution of Eq.~\ref{eq:zer} is the 
singularity of the Jacobian matrix, $\partial \vec \calZ (t+\Delta t)/ \partial 
\vec Y (t+ \Delta t)$.
From Eq.~\ref{eq:zer}, the individual matrix elements of the Jacobian have 
the form
\begin{eqnarray}
	 {\partial \calZ_{i} \over \partial Y_{j}} & = & 
	{\delta_{ij} \over \Delta t} - {\partial \dot Y_{i} \over 
	\partial Y_{j}}\nonumber  \\
	& = & {\delta_{ij} \over \Delta t} - \sum {1 \over \tau_{j}(i)} \ ,
	\label{eq:jacobian}
\end{eqnarray}
where $\delta_{ij}$ is the Kronecker delta, and $\tau_{j}(i)$ is the 
destruction timescale of nucleus $i$ with respect to nucleus $j$ for a 
given reaction, as defined in Eq.~\ref{eq:tau}.  The sum accounts for the 
fact that there may be more than one reaction by which nucleus $j$ is 
involved in the creation or destruction of nucleus $i$.  Along the diagonal 
of the Jacobian, there are two competing terms, $1/\Delta t$ and $\sum 
1/\tau_{i}(i)$.  This sum is over all reactions which destroy nucleus $i$, 
and is dominated by the fastest reactions.  As a result, $\sum 1/\tau_{i}(i)$ 
can be orders of magnitude larger than the reciprocal of the desired timestep, 
$1/\Delta t$.  This is especially a problem near equilibrium, 
where both destruction and the balancing production timescales are very 
short in comparison to the preferred timestep size, resulting in 
differences close to the numerical accuracy (i.e. 14 or more orders of 
magnitude).  In such cases, the term $1/\Delta t$ is numerically neglected, 
leading to numerically singular matrices.  One approach to avoiding this 
problem is to artificially scale these short, equilibrium timescales by a 
factor which brings their timescale closer to $\Delta t$, but leaves them 
small enough to ensure equilibrium.  While this approach has been used 
successfully, the ad hoc nature of this artificial scaling renders these 
methods fragile.  A more promising approach is to make directly use of 
equilibrium expressions for abundances, which, as we will discuss in \S
\ref{sect:hyb}, also assures the economical use of computer resources.

\subsection{Taking Advantage of Matrix Sparseness}\label{sect:sparse}

For larger networks, the Newton-Raphson method requires solution of a 
moderately large ($N=100-3000$) matrix equation.  Since general solution of 
a dense matrix scales as $O(N^{3})$, this can make these large networks 
progressively much more expensive.  While in principal, every species reacts 
with each of the hundreds of others, resulting in a dense Jacobian matrix, 
in practice it is possible to neglect most of these reactions.  Because of 
the $Z_{i}Z_{j}$ dependence of the repulsive Coulomb term in the nuclear 
potential, captures of free neutrons and isotopes of H and He on heavy 
nuclei occur much faster than fusions of heavier nuclei.  Furthermore, with 
the exception of the Big Bang nucleosynthesis and PP-chains, reactions 
involving secondary isotopes of H (deuterium and tritium) and He are 
neglectable.  Likewise, photodisintegrations tend to eject free nucleons or 
\alp-particles.  Thus, with a few important exceptions, for each nucleus we 
need only consider twelve reactions linking it to its nuclear neighbors by the 
capture of an $n,p,\alpha$ or $\gamma$ and release a different one of these 
four.  The exceptions to this rule are the few heavy ion reactions important 
for burning stages like carbon and oxygen burning where the dearth of light 
nuclei cause the heavy ion collisions to dominate.

\begin{figure}
	\centering
 	\includegraphics[width=\textwidth]{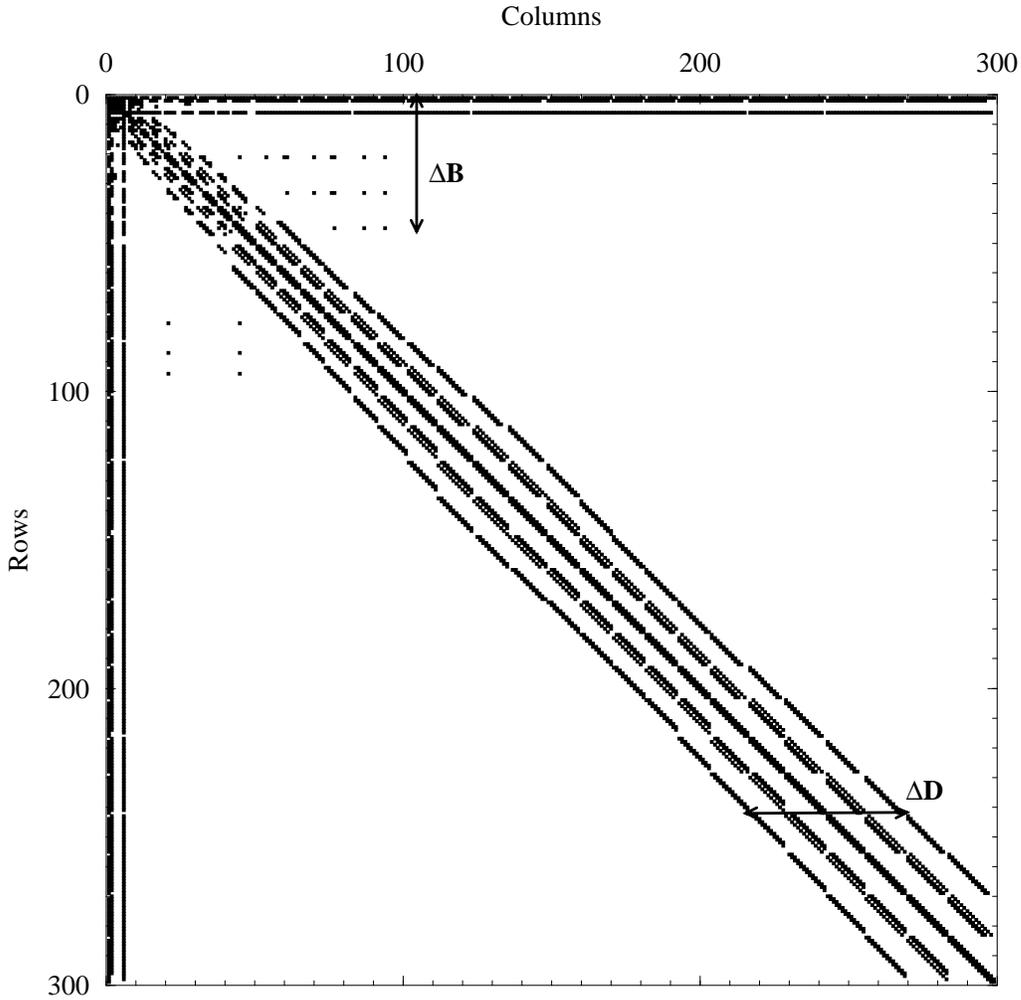}
	\caption{Graphic demonstration of the sparseness of the Jacobian
	matrix.  The filled squares represent the non-zero elements.}
\label{fig:sparse} 
\end{figure}

Fig.~\ref{fig:sparse} demonstrates the sparseness of the resulting Jacobian 
matrix, for a 300 nuclei network designed for silicon burning, but
capable of handling all prior burning stages.  Of
the 90,000 matrix elements, less than 5,000 are non-zero.  In terms of
the standard forms for sparse matrices, this Jacobian is best described
as doubly bordered, band diagonal.  With a border width, $\Delta B$, of
45 necessary to include the heavy ion reactions among \mnuc{C}{12}, 
\mnuc{O}{16} and \mnuc{Ne}{20} along with the free neutrons, protons and
\alp -particles and a band diagonal width, $\Delta D$, of
54, even this sparse form includes almost 50,000 elements.  With solution
of the matrix equation consuming 90+\% of the computational time, there
is clearly a need for custom tailored solvers which take better advantage 
of the sparseness of the Jacobian.  To date best results for small (N$<$100)
matrices are obtained with machine optimized dense solvers (\eg\ LAPACK) or
matrix specific solvers generated by symbolic processing \cite{Muel86,Muel98}.
For large matrices, generalized sparse solvers, both custom built and from
software libraries, are used (see, \eg, \cite{Timm99}).  

\subsection{Physically Motivated Network Specialization}

Often from a physical understanding one can specialize the general solution
method and thereby greatly reduce the computational cost.  As an example of 
such, we will in this section discuss the r-process approximation of 
\cite{CoTT91} (see also \cite{CoCT83,FRRK99}).  For nuclei with $A > 100$, 
charged particle captures (proton and \alp) as well as their reverse
photodisintegrations virtually cease when $T<3\gk$.  This leaves only neutron 
captures and their reverse photodisintegration reactions, as well as 
$\beta$-decays, which can also lead to the emission of delayed neutrons (here
we consider the release of up to three delayed neutrons).  In this case,
Eq.~\ref{eq:ydot} greatly simplifies, leaving 
\begin{eqnarray} 
\dot{Y}(\mnuc{Z}{A})  & = &n_n \langle \sigma v
\rangle^{n,\gamma}_{Z,A-1} Y(\mnuc{Z}{A-1})  + \lambda^{\gamma}_{Z,A+1}
Y(\mnuc{Z}{A+1})  + \sum_{j=0}^3 \lambda_{Z-1,A+j}^{\beta j n}
Y(\mnuc{Z-1}{A+j}) \nonumber \\ & - & \left(n_n \langle \sigma v
\rangle^{n,\gamma}_{Z,A}+\lambda^{\gamma}_{Z,A}  + \sum_{j=0}^3
\lambda_{Z,A}^{\beta j n} \right) Y(\mnuc{Z}{A}) \ , \label{eq:rproc}
\end{eqnarray} 
where $\langle \sigma v \rangle^{n,\gamma}_{Z,A}$ and $\lambda^{\gamma}_{Z,A}$ 
are the velocity integrated neutron capture cross section and the
photodisintegration rate for the nucleus \mnuc{Z}{A}, while
$\lambda_{(Z,A)}^{\beta j n}$ is the decay constant for the $\beta^-$ decay of
\mnuc{Z}{A}, with $j$ delayed neutrons.  The assumption is made that the
neutron abundance $(Y_n=n_n/\rho N_A)$varies slowly enough that it may be evolved explicitly.  One
can see that in Eq.~\ref{eq:rproc}, with $n_n$ thereby fixed, the time
derivatives of each species have a linear dependence on only the abundances of
their neighbors in the same isotopic chain (nuclei with the same $Z$), or that
with one less proton $(Z-1)$. One can then divide the network into separate
pieces for each isotopic chain, and solve them sequentially, beginning with the
lowest Z.  The ``boundary'' terms for this lowest $Z$ chain can be supplied by
a previously  run or concurrently running full network calculation which need
extend only  to this $Z$.  This reduces the solution of a matrix with more
than a thousand rows to the  solution of roughly 30 smaller matrices. 
Furthermore each of these smaller  matrices is also tridiagonal increasing
speed further.  Freiburghaus \etal\ \cite{FRRK99} tested  the assumption of slow variation in the
neutron abundance, and have demonstrated  the usefulness of this method in
r-process simulations, achieving a large decrease in computational cost. 
A similar treatment has been successfully applied to explosive hydrogen burning
based on the assumption of slowly varying proton and alpha abundances
\cite{RFRT97}.  As we will discuss in  \S \ref{sect:hyb}, for other burning 
stages there exist physically motivated simplifications to the general network 
solution method.

\section{Equilibria in Nuclear Astrophysics} \label{sect:nse}

As is the case in many disciplines, equilibrium expressions are frequently 
employed to simplify nuclear abundance calculations.  In most such cases of 
interest in nuclear astrophysics, the fast strong and electromagnetic 
reactions reach equilibrium while those involving the weak nuclear force do 
not.  Since the weak reactions are not equilibrated, the resulting {\em 
Nuclear Statistical Equilibrium} (NSE) requires monitoring of weak 
reaction activity.  Even with this stricture, NSE offers many advantages, 
since hundreds of abundances are uniquely defined by the thermodynamic 
conditions and a single measure of the weak interaction history or the 
degree of neutronization.  Computationally, this reduction in the number of 
independent variables greatly reduces the cost of nuclear abundance evolution.  
Because there are fewer variables to follow within a hydrodynamic model, the 
memory footprint of the nuclear abundances is also reduced, an issue of 
importance in modern multi-dimensional models of supernovae.  Finally, the 
equilibrium abundance calculations depend on binding energies and partition 
functions, quantities which are better known than many reaction rates.  
This is particularly true for unstable nuclei and for conditions where the 
mass density approaches that of the nucleus itself, resulting in exotic 
nuclear structures.

The expression for NSE is commonly derived using either chemical potentials 
or detailed balance (see, \eg, \cite{Clay83,ClTa65,Came79,HaWE85,HTFT99}).  
For a nucleus $^AZ$, composed of $Z$ protons and $N=(A-Z)$ neutrons, in 
equilibrium with these free nucleons, the chemical potential of $^AZ$ can 
be expressed in terms of the chemical potentials of the free nucleons
\begin{equation}
    \mu_{Z,A} = Z \mu_p + N \mu_n \ .
    \label{eq:muAZ}
\end{equation}
For a collection of particles obeying Boltzmann statistics, the chemical 
potential, including rest mass, of each species is given by
\begin{equation}
    \mu_i = m_i c^2 + k_B T \ln \left[\rho N_A{Y_i \over G_i} \left( 
    2\pi\hbar^2 \over {m_i k_B T} \right)^{3 \over 2} \right]
    \label{eq:mu}
\end{equation}
(\eg, \cite{LaLi58}).  Substituting Eq.~\ref{eq:mu} into Eq.~\ref{eq:muAZ} allows 
derivation of an expression for the abundance of every nuclear species in 
terms of the abundances of the free protons ($Y_p$) and neutrons ($Y_n$),
\begin{eqnarray}
    Y(^AZ) &=& {G(^AZ) \over 2^A} {\left(\rho N_A \over \theta \right)}^{A-1} 
    A^{3 \over 2} \exp {\left( B(^AZ) \over {k_B T} \right)} {Y_n}^N {Y_p}^Z 
	\nonumber \\ 
	&\equiv& C(^AZ) {Y_n}^N {Y_p}^Z \ , 
    \label{eq:nse}
\end{eqnarray}
where $G(^AZ)$ and $B(^AZ)$ are the partition function and binding energy 
of the nucleus $^AZ$, $N_A$ is Avagadro's number, $k_B$ is Boltzmann's 
constant, $\rho$ and $T$ are the density and temperature of the plasma, and 
$\theta$ is given by
\begin{equation}
    \theta = \left( {m_u k_B T \over 2 \pi \hbar^2 } \right)^{3/2} \ .
    \label{eq:theta}
\end{equation}

\begin{figure}
	\centering
 	\includegraphics[angle=90,width=\textwidth]{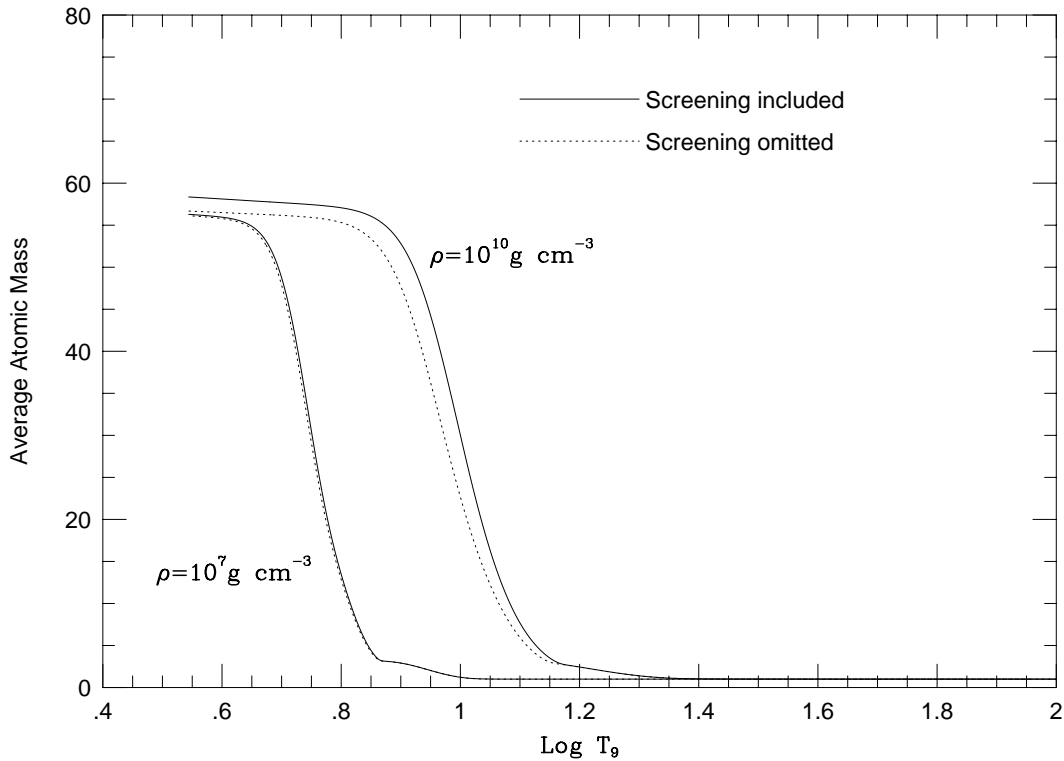}
	\caption{The average atomic mass for material in NSE as a function of 
	Temperature.  The solid lines include screening corrections to the 
	nuclear binding energies, while the dotted lines ignore this effect.  See 
	\cite{HTFT99} for more discussion of the importance of screening in NSE.} 
\label{fig:AbarNSE} 
\end{figure}

Thus abundances of all nuclear species can be expressed as functions of 
two.  Mass conservation $(\sum A Y = 1)$ provides one constraint.  The 
second constraint is the amount of weak reaction activity, often expressed 
in terms of the total proton abundance, $\sum Z Y$, which charge 
conservation requires equal the electron abundance, \ye.  Thus the nuclear 
abundances are uniquely determined for a given $(T,\rho,\ye)$.  Alternately, 
the weak interaction history is sometimes expressed in terms of the 
neutron excess $\eta=\sum(N-Z) Y$.  Figure~\ref{fig:AbarNSE} displays the 
temperature and density dependence of $\bar A = \sum A Y / \sum Y = 1/ \sum 
Y$, the average nuclear mass of the NSE distribution.  At high 
temperatures, free nucleons are favored, hence $\bar A \sim 1$.  For 
intermediate temperatures the compromise of retaining large numbers of 
particles while increasing binding energy favors \mnuc{He}{4}, which has 
$80\%$ of the binding energy of the iron peak nuclei.  At low temperatures, 
Eq.~\ref{eq:nse} strongly favors the most bound nuclei, the iron peak 
nuclei, so $\bar A \rightarrow 60$ as the temperature drops.  Density can 
be seen to scale the placement of these divisions between high, 
intermediate and low temperature.  Variations in \ye\ do not strongly 
affect Figure~\ref{fig:AbarNSE}.  At high temperatures, it simply effects 
the ratio of $Y_{p}/Y_{n} \approx \ye/1-\ye$.  At low temperatures, 
variation in \ye\ changes which Fe-peak isotopes dominate.  Though 
\mnuc{Ni}{56} is less tightly bound than \mnuc{Fe}{54}, it is more tightly 
bound than \mnuc{Fe}{54} + 2 \mnuc{H}{1}, which would be required by charge 
conservation if $\ye\sim .5$.  Thus $Y(\mnuc{Ni}{56}) > Y(\mnuc{Fe}{54})$ 
for low $T$ with $\ye\sim .50$, but $Y(\mnuc{Fe}{54}) > Y(\mnuc{Ni}{56})$ 
for smaller \ye.  In general, the most abundant nuclei at low temperatures 
are the most bound nuclei for which $Z/A \sim \ye$.

As with any equilibrium distribution, there are limitations on the 
applicability of NSE. The first requirement for NSE to provide a good 
estimate of the nuclear abundances is that the temperature be sufficient 
for the endoergic reaction of each reaction pair to occur.  Since for all 
particle-stable nuclei between the proton and neutron drip lines (with the 
exception of nuclei unstable against alpha decay), the photodisintegrations 
are endoergic, with typical Q-values among ($\beta$) stable nuclei of 8-12 
\mev, by Eq.~\ref{eq:detbal} this requirement reduces to $T>3\gk$.  While 
this requirement is necessary, it is not sufficient.  In the case of 
hydrostatic silicon burning, even when this condition is met, appreciable 
time is required to convert Si to Fe-peak elements.  In the case of 
explosive silicon burning, the adiabatic cooling on timescales of seconds 
can cause conditions to change more rapidly than NSE can follow, breaking 
NSE down first between \mnuc{He}{4} and \mnuc{C}{12}, at $T\sim 6\gk$ 
\cite{MeKC98} and later between the species near silicon and the Fe-peak 
nuclei, at $T\sim 4\gk$ \cite{HiTh99}.  Thus it is clear that in the face 
of sufficiently rapid thermodynamic variations, NSE provides a problematic 
estimate of abundances.

\section{Merging Equilibria with Nuclear Networks} \label{sect:hyb}

In spite of the limitations on the applicability of NSE, the reduced 
computational cost provides a strong motivation to maximize the use of 
equilibria.  The use of equilibrium expressions for single abundances is, 
in fact, common in nuclear reaction networks, typically to track the 
abundances of short-lived unstable intermediates in ``three-particle'' 
processes.  The most common example of this is the triple \alp\ process, 
\begin{equation}
    \begin{array}{rcllll}
         \mnuc{He}{4} + \mnuc{He}{4} &\rightleftharpoons&\mnuc{Be}{8}&
	 & & Q = -.09 \mev \\
         \mnuc{He}{4} + \mnuc{Be}{8} &\rightleftharpoons&\mnuc{C^*}{12}
         &\rightarrow & \mnuc{C}{12} + \gamma \qquad & Q = 7.37 \mev \ ,
    \end{array}
\end{equation} 
by which Helium burning occurs.  With $\tau(\mnuc{Be}{8}) \sim 10^{-16} \sec$, 
only rarely does a \mnuc{Be}{8} survive long enough for a second \alp\ to 
capture.  As a result of the near balance of the first reaction pair, the 
abundance of \mnuc{Be}{8} can be expressed in terms of the \alp-particle 
abundance,
\begin{equation}
Y(\mnuc{Be}{8})= {\rho N_A \over \theta} \left(1 \over 2\right) ^{3/2}
\exp \left(M(\mnuc{Be}{8})-2M_{\alp}\over {k_B T}\right) Y_{\alp}^2 .
\end{equation}
     
Likewise for temperatures in excess of .1 \gk, the most likely result following
the second \alp\ capture to form an excited state of \mnuc{C}{12} is a decay
back to \mnuc{Be}{8} ($\Gamma_{\alp}(\mnuc{C^*}{12})/\Gamma_{\gamma}
(\mnuc{C^*}{12})>10^3$), thus the abundance of \mnuc{C^*}{12} is well
characterized, via $\mu(\mnuc{C^*}{12}) = 3 \mu(\mnuc{He}{4})$, by
\begin{equation}
Y(\mnuc{C^*}{12})= \left(\rho N_A \over \theta \right)^2 \left(3 \over 
16\right)^{3/2} \exp \left(M(\mnuc{C^*}{12})-3M_{\alp}\over {k_B T}\right) 
Y_{\alp}^3 .
\end{equation}
When this is the case, the effective triple \alp\ reaction rate is
simply that of the decay of \mnuc{C}{12} from the excited state to the
ground state,
\begin{equation}
r_{3\alp} = \rho N_A Y(\mnuc{C*}{12}) \Gamma_{\gamma}(\mnuc{C^*}{12}) / \hbar \ .
\end{equation}
 
This use of local equilibrium within a rate equation shares many
characteristics with the more elaborate schemes we will discuss later in this
section. The number of species tracked by the network is reduced since
$Y(\mnuc{Be}{8})$ need not be directly evolved.  Problematically small timescales
like $\tau(\mnuc{Be}{8})$ are removed, replaced by larger time scales
($\tau_{3\alp} \sim 10^5-10^7$ years during core helium burning).  The
non-linearity of network time derivatives is increased ($\dot Y(\mnuc{C}{12})
\propto Y_{\alp}^3$) under this scheme.  This approximation also breaks down 
at low $T$ (for details see \cite{NoTM85}). 

In addition to silicon burning, there  are a number of astrophysically
important situations where $T > Q/30k_{B}$  for at least some of the relevant
reactions and so large equilibrium  groups exist, but NSE is not globally
valid.  This include the r-process \cite{CoTT91,Krea93,Boea96} and the
rp-process in novae and X-ray bursts \cite{RFRT97,SAGW98}, where neutron or
proton separation energies ($Q_n$ or $Q_p$) of 2~\mev and less are  often
encountered.  Beginning with \cite{BoCF68}, a number of attempts have  been
made to take advantage of these partial equilibria to reduce the  number of
independent variables evolved via rate equations and thereby  reduce the
computational cost of modeling these burning stages.  Since we  lack the time
to discuss all in detail, we refer the reader to  \cite{FKRT97} and
\cite{RFRT97} for discussion of hybrid networks for the  r-process and
rp-process, respectively.  Instead, we will here concentrate  on the
application of hybrid equilibria networks to silicon burning 
\cite{HiTh96,MeKC98,HiTh99,HiFT99}.  In the present context we will concentrate
on the QSE-reduced \alp -network \cite{HKWT98}, a simple, but pedagogically 
illustrative, example which details  the main ideas behind such hybrid
equilibrium networks.  Quasi-equilibrium  (QSE) is a term coined by Bodansky,
Clayton \& Fowler \cite{BoCF68} to  describe the local equilibrium groups which
form during silicon burning.

\subsection {The QSE-reduced \alp-network}\label{sect:qse}

Tracking the nuclear evolution during the major energy producing burning 
stages from the exhaustion of hydrogen through to the establishment of NSE 
requires, at minimum, a network that includes nuclei from \alp-particles to 
Zn.  As we discussed in \S \ref{sect:astro}, silicon 
burning presents a particular problem as material proceeds from silicon to 
the iron peak not via heavy ion captures but through a chain of 
photodisintegrations and light particle captures.  We will discuss here the 
minimal nuclear set which can follow this evolution, the set of 
\alp-particle nuclei; \alp, \mnuc{C}{12}, \mnuc{O}{16}, \mnuc{Ne}{20}, 
\mnuc{Mg}{24}, \mnuc{Si}{28}, \mnuc{S}{32}, \mnuc{Ar}{36}, \mnuc{Ca}{40}, 
\mnuc{Ti}{44}, \mnuc{Cr}{48}, \mnuc{Fe}{52}, \mnuc{Ni}{56}, \mnuc{Zn}{60}.  
For convenience we will label this full set \calF\ and refer to its 
abundances as \yf.  Silicon burning in fact presents a larger problem, as 
the nuclear flow from silicon to the iron peak nuclei does not generally 
proceed through nuclei with N=Z, especially when significant neutronization 
has occurred \cite{HiTh96}.  In some hydrodynamical models, however, such 
compromise is made necessary by the computational limitations, either the 
time necessary to solve larger networks or the hydrodynamical problems 
associated with evolving and storing a large number of abundances.  
Furthermore, the small size of the \alp\ network (14 nuclei and 17 
reactions) makes application of QSE to \alp-chain nucleosynthesis a 
pedagogically useful example.

The objective of the QSE-reduced \alp-network is to evolve \yf\ (and 
calculate the resulting energy generation) in a more efficient way.  Under 
conditions where QSE applies, the existence of the silicon and iron peak 
QSE groups (which are separated by the nuclear shell closures Z=N=20 and 
the resulting small Q-values and reaction rates) allows calculation of 
these 14 abundances from 7.  For the members of the silicon group 
(\mnuc{Si}{28}, \mnuc{S}{32}, \mnuc{Ar}{36}, \mnuc{Ca}{40}, \mnuc{Ti}{44}) 
and the iron peak group (\mnuc{Cr}{48}, \mnuc{Fe}{52}, \mnuc{Ni}{56}, 
\mnuc{Zn}{60}) the individual abundances can be calculated by expressions 
similar to Eq.~\ref{eq:nse},
\begin{eqnarray}
    Y_{QSE,\mathrm{Si}}(^AZ) & = & {{C(^AZ)}\over{C(\mnuc{Si}{28})}} 
    Y(\mnuc{Si}{28}) Y_{\alpha}^{{A-28} \over 4} \nonumber  \\
    Y_{QSE,\mathrm{Ni}}(^AZ) & =  & {{C(^AZ)}\over{C(\mnuc{Ni}{56})}} 
    Y(\mnuc{Ni}{56}) Y_{\alpha}^{{A-56} \over 4},
    \label{eq:yq}
\end{eqnarray}
where ${C(^AZ)}$ is defined in Eq.~\ref{eq:nse} and $(A-28)/4$ and 
$(A-56)/4$ are the number of \alp-particles needed to construct $^{A}Z$ 
from \mnuc{Si}{28} and \mnuc{Ni}{56}, respectively.  Thus, where QSE 
applies, \yf\ is a function of \yr, where the reduced nuclear set \calR\ is 
defined as \alp, \mnuc{C}{12}, \mnuc{O}{16}, \mnuc{Ne}{20}, \mnuc{Mg}{24}, 
\mnuc{Si}{28}, \mnuc{Ni}{56}, and we need only evolve \yr.  It should be 
noted that \mnuc{Mg}{24} is ordinarily a member of the silicon QSE group 
\cite{Arne96,WoAC73,HiTh96}, but for easier integration of 
prior burning stages with a conventional nuclear network, we will evolve 
\mnuc{Mg}{24} independently.  The main task when applying such hybrid 
schemes is finding the boundaries of QSE groups and where individual 
nuclei have to be used instead.  Treating marginal group members as part of 
a group increases the efficiency of the calculation, but may decease the 
accuracy.

While \yr\ is a convenient set of abundances for calculating \yf, it is not 
the most efficient set to evolve, primarily because of the non-linear 
dependence on $Y_{\alpha}$.  Instead we define \yg = $[Y_{\alpha G}$, 
$Y(\mnuc{C}{12})$, $Y(\mnuc{O}{16})$, $Y(\mnuc{Ne}{20})$, $Y(\mnuc{Mg}{24})$, 
$Y_{SiG}$, $Y_{FeG}]$ where
\begin{eqnarray}
    Y_{\alp G} & = & \quad Y_{\alpha} + \sum_{i \in Si \ group} 
	{{A_{i}-28} \over 4} Y_{i} + \sum_{i \in Fe \ group} {{A_{i}-56} \over 
	4} Y_{i} \ , \nonumber \\
    Y_{Si G} & = & \sum_{i \in Si \ group} Y_{i} \ , \label{eq:yg} \\
    Y_{Fe G} & = & \sum_{i \in Fe \ group} Y_{i} \ . \nonumber 
\end{eqnarray}
Physically, $Y_{\alpha G}$ represents the sum of the abundances of free 
\alp-particles and those \alp-particles required to build the members of the 
QSE groups from \mnuc{Si}{28} or \mnuc{Ni}{56}, while $Y_{Si G}$ and $Y_{Fe
G}$  represent the total abundances of the silicon and iron peak QSE groups.
This method, which here is applied only to the chain of \alp -nuclei can also
be generalized to arbitrary networks \cite{HiFT99}.  For larger networks 
which contain nuclei with $N\neq Z$, one must be able to follow the abundances
of free neutrons and protons, particularly since weak interactions will change
the global ratio of neutrons to protons.  In place of $Y_{\alpha G}$ in
Eq.~\ref{eq:yg}, one constructs  $Y_{NG}= \sum_{i,light} N_i Y_i+ \sum_{i,Si}
(N_i-14) Y_i +  \sum_{i,Fe} (N_i-28) Y_i$ and $Y_{ZG}= \sum_{i,light} Z_i Y_i
+ \sum_{i,Si} (Z_i-14) Y_i + \sum_{i,Fe} (Z_i-28) Y_i $, if \mnuc{Si}{28} and
\mnuc{Ni}{56} are chosen as the focal nuclei for the Si and Fe groups.  

Corresponding to this reduced set of abundances \calG\ is a reduced set of 
reactions, with quasi-equilibrium allowing one to ignore the reactions among 
the members of the QSE groups. Unfortunately, the rates of these remaining 
reactions are functions of the full abundance set, \yf, and are not easily 
expressed in terms of the group abundances, \yg.  Thus, for each \yg, one must 
solve for \yr\ and, by  Eq.~\ref{eq:yq}, \yf, in order to calculate \ygdot\
which is needed to evolve \yg\ via Eq.~\ref{eq:deriv}.  Furthermore,
Eq.~\ref{eq:dely} requires the calculation of the Jacobian of $\vec \calZ$, 
which can not be calculated directly since \ygdot\ cannot be expressed in 
terms of \yg.  Instead we find it sufficient to use the chain rule,
\begin{equation}
	{{\partial \ygdot} \over {\partial \yg}} = {{\partial \ygdot} \over 
	{\partial \yr}} {{\partial \yr} \over {\partial \yg}}	
	\label{eq:chain}
\end{equation}
to calculate the Jacobian.  Analytically, the first term of the chain rule 
product is easily calculated from the sums of reaction terms, while the 
second term requires implicit differentiation using Eq.~\ref{eq:yg}, which is
discussed further by Hix \etal\ \cite{HKWT98}.  

\subsection{Silicon burning with the QSE-reduced \alp\ network}

\begin{figure}
    \includegraphics[angle=90,width=\textwidth]{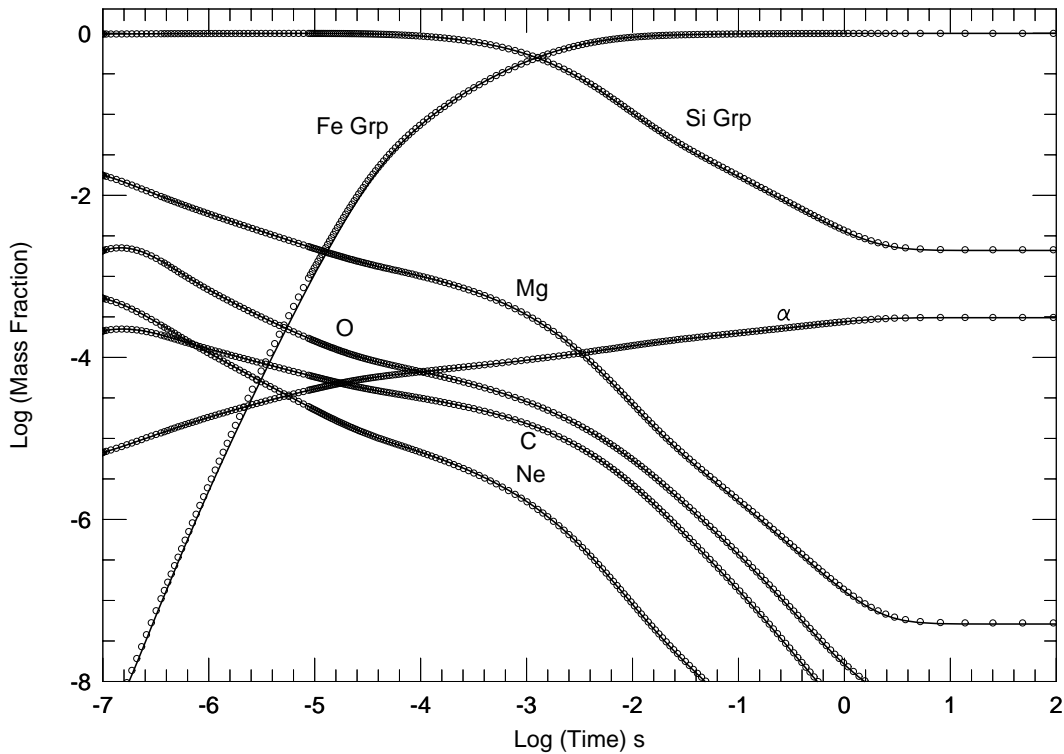}
    \caption{Evolution of the independent nuclear mass fractions for 
    constant thermodynamic conditions, $T=5 \gk$ and $\rho=10^{9} \gcc$.  
    The solid lines display the evolution due to a conventional \alp 
    -network, the circles show the evolution by the QSE-reduced \alp\ network.}
    \label{fig:xc59} 
\end{figure}

In this section we will demonstrate the accuracy with which the QSE-reduced 
\alp\ network duplicates the results of the full 14 element \alp\ network 
for silicon burning.  Our first example is a nucleosynthesis calculation 
occurring under constant temperature and density.  While such calculations 
provide the least challenging comparison, they also allow comparison with 
NSE, which should represent the final abundances for these calculations.  
Figure~\ref{fig:xc59} offers comparison of the mass fractions of the 7 
independent species; \alp-particles, \mnuc{C}{12}, \mnuc{O}{16}, 
\mnuc{Ne}{20}, \mnuc{Mg}{24}, and the silicon and iron peak groups, as 
evolved by the QSE-reduced and conventional \alp\ networks for silicon 
burning at $5 \gk$ and a density of $10^{9} \gcc$.  Apart from an early 
enhancement by the QSE-reduced network of the iron peak mass fraction (20\% 
after $10^{-6}$ seconds), these mass fractions typically agree to within 
1\%.  Since the nuclear energy release depends linearly on the abundance 
changes (see Eq.~\ref{eq:edot}), differences in small abundances have little 
effect on the nuclear energy store.  In this case, the difference in the 
rate of energy generation calculated by the two networks is $<1\%$ at 
$10^{-6}$ and $10^{-4}$ seconds.  This difference is significantly smaller 
than the variation, shown by both networks, in the rate of energy 
generation between timesteps, with \edot\ typically declining by 5\% per 
timestep over this interval.

Significant variations in abundance among the individual members of the QSE 
groups between the two networks are also limited to the small abundances.  At 
early times, the small abundances within the iron peak reduce the accuracy 
of QSE at predicting the individual abundances of members of the iron peak 
group.  Much of the enhanced mass fraction of the iron peak nuclei at early 
times, seen in Fig.~\ref{fig:xc59}, is due to the QSE-reduced network's 
emphasis on heavier nuclei at the expense of \mnuc{Cr}{48}.  After an 
elapsed time of $10^{-6}$ seconds, the average mass of the iron peak 
nuclei, $\bar A_{FeG}$, is 49.2 according to the conventional network and 
52.6 according to the QSE-reduced network.  As a result, the abundances of 
\mnuc{Cr}{48} and \mnuc{Fe}{52} calculated by the QSE-reduced network are 
38\% and 164\% of their conventional network values, while \mnuc{Ni}{56} 
and \mnuc{Zn}{60} are 16 times more abundant than the conventional network 
predicts.  However, the total Fe group mass fraction is only $10^{-7}$ at 
this point in time.  As the iron peak nuclei become more abundant, QSE 
provides a better estimate of the relative abundances within the group, 
reducing such discrepancies.  By the time the iron peak nuclei represent a 
significant portion of the mass, the differences in the abundance 
predictions for all nuclei are only a few per cent.  As each network 
reaches its respective equilibrium, after an elapsed time of 24 seconds, the 
abundance predictions of these networks (shown in Table~\ref{tab:const1}) 
differ by at most 3\%, even among the nuclei with the smallest abundances.  
Not surprisingly, in view of these small abundance differences, the 
difference in the total energy released by these networks is less than 
.1\%.  Comparison of the network abundances with abundances calculated from 
NSE show a similarly low level of difference.  

\begin{table}
    \centering
    \caption{Comparison of equilibria calculated by the conventional and 
    QSE-reduced \alp -networks with NSE for $T=5 \gk$ and $\rho=10^9 \gcc$}
    \label{tab:const1}
    \begin{tabular}{clll}
    	\hline
    	Nucleus & \multicolumn{1}{c}{$Y_{net}$}&\multicolumn{1}{c}{$Y_{qse}$} 
    	& \multicolumn{1}{c}{$Y_{nse}$} \\
    	\hline
    	\mnuc{He}{4} & 7.73\ttt{-5} & 7.78\ttt{-5} & 7.80\ttt{-5}  \\ 
    	\mnuc{C}{12} & 1.01\ttt{-10}& 1.02\ttt{-10}& 1.01\ttt{-10} \\
    	\mnuc{O}{16} & 3.23\ttt{-10}& 3.30\ttt{-10}& 3.24\ttt{-10} \\
    	\mnuc{Ne}{20}& 4.08\ttt{-12}& 4.19\ttt{-12}& 4.11\ttt{-12} \\
    	\mnuc{Mg}{24}& 2.13\ttt{-9} & 2.20\ttt{-9} & 2.16\ttt{-9}  \\
    	\mnuc{Si}{28}& 4.51\ttt{-6} & 4.68\ttt{-6} & 4.53\ttt{-6}  \\
    	\mnuc{S}{32} & 9.34\ttt{-6} & 9.63\ttt{-6} & 9.37\ttt{-6}  \\
    	\mnuc{Ar}{36}& 1.06\ttt{-5} & 1.09\ttt{-5} & 1.07\ttt{-5}  \\
    	\mnuc{Ca}{40}& 3.01\ttt{-5} & 3.06\ttt{-5} & 3.03\ttt{-5}  \\
       	\mnuc{Ti}{44}& 1.68\ttt{-6} & 1.70\ttt{-6} & 1.69\ttt{-6}  \\
    	\mnuc{Cr}{48}& 3.75\ttt{-5} & 3.80\ttt{-5} & 3.75\ttt{-5}  \\
    	\mnuc{Fe}{52}& 8.64\ttt{-4} & 8.70\ttt{-4} & 8.65\ttt{-4}  \\
    	\mnuc{Ni}{56}& 1.70\ttt{-2} & 1.70\ttt{-2} & 1.70\ttt{-2}  \\
    	\mnuc{Zn}{60}& 4.71\ttt{-6} & 4.68\ttt{-6} & 4.69\ttt{-6}  \\
    	\hline
    \end{tabular}
\end{table}

While example calculations of silicon burning under constant conditions are 
instructive, if the QSE-reduced \alp-network is to replace a conventional 
\alp-network it must be shown to be accurate under changing thermodynamic 
conditions.  Of particular importance is the ability to model explosive 
silicon burning.  Because the products of hydrostatic silicon burning are 
trapped deep in the potential well of their parent star, it is only by 
explosion that the interstellar medium is enriched in intermediate mass and 
iron peak elements.  To model silicon burning occurring as a result of 
shock heating, we will follow the approximation introduced by \cite{FoHo64}. 
Therein a mass zone is instantaneously heated by a passing shock to some 
peak initial temperature, $T_{i}$, and density, $\rho_i$, and then expands 
and cools adiabatically, with the timescale for the expansion given by the 
free fall timescale, $\tau_{\rm HD} = \left(24 \pi G \rho \right)^{-1/2} = 
446 \rho_6^{-1/2} \mathrm{milliseconds}$.

\begin{figure}
 	\includegraphics[angle=90,width=\textwidth]{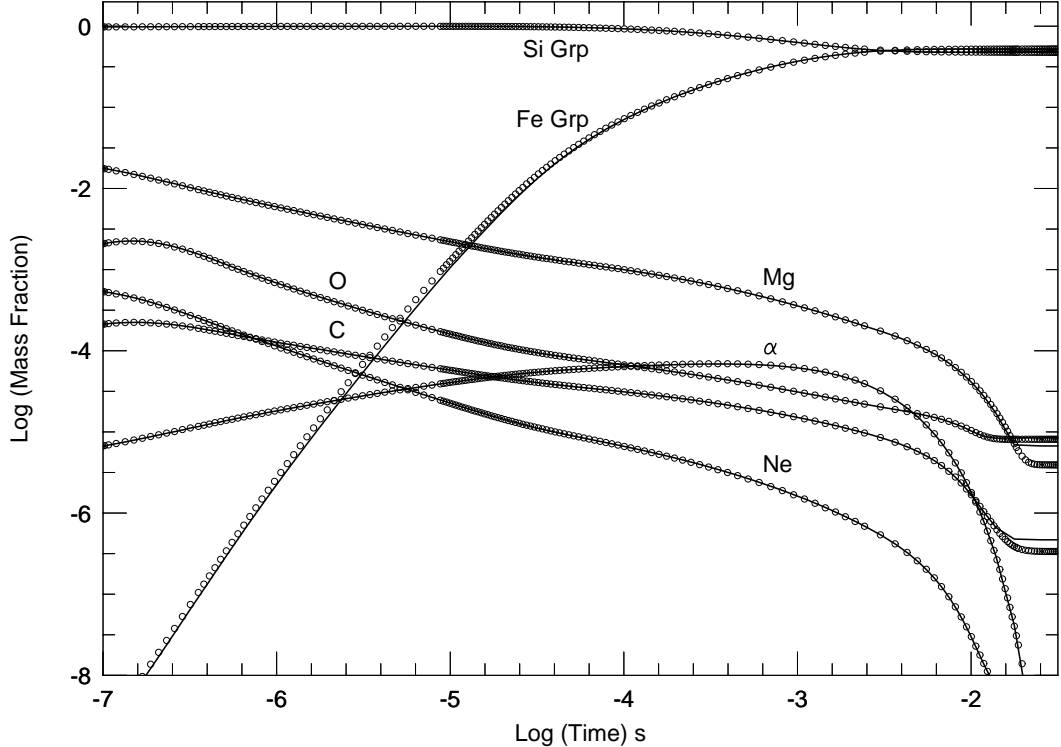}
	\caption{Evolution of the independent nuclear mass fractions under 
	adiabatic expansion with $T_{i}=5 \gk$ and $\rho_{i}= 10^{9} \gcc$.  
	The solid lines display the evolution due to a conventional \alp\ 
	network, the circles show the evolution by the QSE-reduced \alp\ 
	network.}
	\label{fig:xe59} 
\end{figure}

Figure~\ref{fig:xe59} shows the nuclear evolution for an example of this 
explosive burning model with $T_{i}=5 \gk$ and $\rho_{i}=10^9 \gcc$.  Over 
the first millisecond, the evolution portrayed here is virtually identical 
to that of Figure~\ref{fig:xc59}; however by the time one millisecond has 
elapsed, the temperature has dropped to 4.9 \gk, slowing the reactions 
which are turning silicon into iron peak elements.  This cooling, which 
drops $T$ below 4 \gk\ after 9 ms and below 3 \gk\ after 22 ms, freezes out 
the nuclear reactions before NSE is reached, resulting in incomplete 
silicon burning, as discussed by Woosley \etal\ \cite{WoAC73}. In this case, 
the freezeout leaves nearly equal amounts of silicon group and iron peak group 
elements.  Throughout most of the evolution in this example, the agreement 
between the mass fractions as evolved by the QSE-reduced \alp\ network with 
those evolved by its conventional counterpart is comparable to that 
demonstrated under constant thermodynamic conditions.  Columns 2 and 3 of 
Table~\ref{tab:nfreeze} compare the abundances after 9 ms have elapsed, 
with $T$ nearing 4 \gk.  In this case, the largest relative error (5\%) is 
in the abundance of \mnuc{Cr}{48}.  These small differences in abundance 
result in small differences in the accumulated nuclear energy release, 
approximately .5\% to this point.  By this time, adiabatic cooling has 
greatly reduced the rate of energy generation from its peak of more than 
$10^{22} \erggs$ to roughly $10^{17} \erggs$.  Though the absolute 
difference in the rate of energy generation as calculated by the two networks 
has declined from $\sim 10^{19}$ to $10^{16} \erggs$, the relative 
difference has grown to 10\% as $T$ nears 4 \gk.  Fortunately this 
difference is negligible, since nuclear energy release has virtually 
ceased, with $<.2\%$ of the total energy release remaining.

\begin{table}
    \centering
	\caption{Comparison of network abundances for $T_{i}=5.0 \gk$ and 
	\rhoi= $10^9 \gcc$}
    \label{tab:nfreeze}
    \begin{tabular}{clllll}
    	\hline
    	Time (ms)&\multicolumn{2}{c}{8.77}&\multicolumn{2}{c}{17.7}
    	&\multicolumn{1}{c}{255} \\
    	$T (\gk)$& \multicolumn{2}{c}{4.07}&\multicolumn{2}{c}{3.29}
    	&\multicolumn{1}{c}{0.01} \\
    	\hline \hline
    	Nucleus & \multicolumn{1}{c}{$Y_{net}$}
    	&\multicolumn{1}{c}{$Y_{qse}$}& \multicolumn{1}{c}{$Y_{net}$} 
    	&\multicolumn{1}{c}{$Y_{qse}$}& \multicolumn{1}{c}{$Y_{net}$} \\
    	\hline
    	\mnuc{He}{4} & 7.90\ttt{-7} & 7.82\ttt{-7} & 1.04\ttt{-8} & 1.01\ttt{-8} & 1.94\ttt{-14}\\
    	\mnuc{C}{12} & 1.96\ttt{-7} & 1.96\ttt{-7} & 3.99\ttt{-8} & 3.23\ttt{-8} & 3.90\ttt{-8} \\
    	\mnuc{O}{16} & 7.34\ttt{-7} & 7.39\ttt{-7} & 5.26\ttt{-7} & 5.07\ttt{-7} & 5.27\ttt{-7} \\
    	\mnuc{Ne}{20}& 2.63\ttt{-9} & 2.63\ttt{-9} & 1.69\ttt{-10}& 1.48\ttt{-10}& 9.88\ttt{-11}\\
    	\mnuc{Mg}{24}& 2.24\ttt{-6} & 2.26\ttt{-6} & 2.88\ttt{-7} & 2.98\ttt{-7} & 2.80\ttt{-7} \\
    	\mnuc{Si}{28}& 7.65\ttt{-3} & 7.76\ttt{-3} & 7.58\ttt{-3} & 7.86\ttt{-3} & 7.58\ttt{-3} \\
    	\mnuc{S}{32} & 4.93\ttt{-3} & 4.96\ttt{-3} & 5.16\ttt{-3} & 5.15\ttt{-3} & 5.16\ttt{-3} \\
    	\mnuc{Ar}{36}& 1.43\ttt{-3} & 1.42\ttt{-3} & 1.27\ttt{-3} & 1.21\ttt{-3} & 1.27\ttt{-3} \\
    	\mnuc{Ca}{40}& 1.32\ttt{-3} & 1.30\ttt{-3} & 1.32\ttt{-3} & 1.22\ttt{-3} & 1.32\ttt{-3} \\
    	\mnuc{Ti}{44}& 7.07\ttt{-6} & 6.90\ttt{-6} & 1.96\ttt{-6} & 1.72\ttt{-6} & 1.69\ttt{-6} \\
    	\mnuc{Cr}{48}& 5.89\ttt{-5} & 5.58\ttt{-5} & 4.40\ttt{-5} & 1.19\ttt{-5} & 4.40\ttt{-5} \\
    	\mnuc{Fe}{52}& 7.17\ttt{-4} & 6.93\ttt{-4} & 6.33\ttt{-4} & 3.05\ttt{-4} & 6.33\ttt{-4} \\
    	\mnuc{Ni}{56}& 8.63\ttt{-3} & 8.60\ttt{-3} & 8.73\ttt{-3} & 9.04\ttt{-3} & 8.73\ttt{-3} \\
    	\mnuc{Zn}{60}& 6.18\ttt{-8} & 6.06\ttt{-8} & 3.26\ttt{-9} & 3.23\ttt{-9} & 4.38\ttt{-10}\\
    	\hline
    \end{tabular}
\end{table}

Though energetically unimportant, nuclear reactions continue, resulting in 
significant changes in the smaller abundances, a point which is discussed 
in detail by \cite{WoAC73,HiTh99}.  The continued cooling drops $T$ below 3 
\gk, gradually freezing out the photodisintegrations responsible for QSE. 
However this same decline in temperature also reduces the rate of charged 
particle capture reactions, greatly reducing the amount of nucleosynthesis 
which occurs after $T$ drops below $\sim 3.5 \gk$.  The group abundances of 
the silicon and iron peak groups (which account for 99.9\% of the mass), as 
calculated by the QSE-reduced \alp\ network after 18 ms have elapsed 
($T=3.3 \gk$), differ by less than 1\% from those of the conventional \alp\ 
network at the same point in time.  Comparison of Columns 4 and 5 of 
Table~\ref{tab:nfreeze} reveals larger variations among individual 
abundances, most notably, significant under estimation by the QSE-reduced 
network of the \mnuc{Cr}{48} and \mnuc{Fe}{52} abundances, with a small, 
compensatory over estimate of the \mnuc{Ni}{56} abundance.  These 
variations, factors of 2 and 4 for \mnuc{Fe}{52} and \mnuc{Cr}{48}, 
respectively, and 3\% for \mnuc{Ni}{56} signal the breakdown of the iron 
peak QSE group.  With the steep decline in temperature and density, the 
flux upward from \mnuc{Fe}{52} in the conventional network is no longer 
sufficient to provide the reduction in abundance which QSE and the sharply 
declining abundance of free \alp-particles requires.

As the temperature and density continue to decline, so too does the free 
\alp-particle abundance.  Column 6 of Table~\ref{tab:nfreeze} details the 
abundances after 255 ms have elapsed, with $T$ having dropped to .01 \gk, 
and all abundances having frozen out.  Comparison of columns 4 and 6 
reveals that the decline of the free \alp-particle abundance from $10^{-8}$ 
is the largest abundance variation beyond 18 ms.  Since the more abundant 
species have effectively frozen out by the time $T$ approaches 3.5 \gk, 
comparison of columns 5 and 6 reveal that the predictions of the 
QSE-reduced network, frozen near $T=3.5 \gk$, also provide good abundance 
estimates, in spite of the effects of differential freezeout.  Thus, we 
have seen that the QSE-reduced $\alp$ network provides an excellent 
description of energy generation and an accurate account of the abundances, 
down to (reaction) freeze-out temperatures of $\sim 3.5 \gk$.  These 
abundances at $T\sim 3.5 \gk$ provide a very good approximation to the 
final results for the dominant nuclei.  More detailed and accurate 
accounting of smaller abundances would require a switch back to the use of 
the full network below these freeze-out temperatures. 

In this section, we have demonstrated that a QSE-reduced \alp\ network can 
be used as a replacement for full 14 element \alp\ network when modeling 
silicon burning, without significant errors in energy generation or 
nucleosynthesis.  For such a small system, the computational benefits of 
such equilibrium network hybrids are also small, a factor of 2 in network 
computational time (due in most part to the factor of 2 reduction in matrix 
size) and a factor of 2 reduction in the number of nuclear variables which 
must be evolved within a hydrodynamic model.  For larger nuclear networks, 
like the realistic hybrid networks for the r-process \cite{FKRT97}, 
rp-process \cite{RFRT97} and silicon burning \cite{HiFT99}, the potential 
for improvement in speed and size is even greater.  In comparison to full 
networks with hundreds of nuclei, reduction in the number of independent 
nuclei by a factor of 2-4 can result in increases in network speed of a 
factor of 5-10 because of the nonlinear relation between matrix size and 
the length of time to solve a matrix equation.  Equally important is the 
reduction in the number of nuclear abundances that are hydrodynamically 
evolved, significantly reducing the memory footprint of such calculations.

\section{Conclusion}

While it is true that our basic picture of the origin of the elements seems 
well validated, a great deal of refinement and investigation remains.  Due 
to the extreme conditions required for thermonuclear reactions, nuclear 
processes in astrophysics most frequently occur in deep gravitational 
potential wells, often strongly obscured from view.  As a result, the 
number of direct observables are few, and nucleosynthetic observables 
depend strongly on the evolution of the stars and stellar explosions which 
produce them.  As each of these related 
fields becomes more sophisticated, greater demands of speed and accuracy 
are placed on nuclear astrophysics calculations, mandating continued 
improvements in the numerical methods used.  The best example of this is 
the recent trend toward multi-dimensional hydrodynamic simulations, which 
greatly increases the number of independent nucleosynthesis calculations 
which must be performed.

In this article, we have concentrated on the numerical methods of use for
nucleosynthesis, describing the important means by which nuclear 
compositions are currently evolved within an astrophysical simulation, 
thermonuclear rate equations and Nuclear Statistical Equilibrium.  While 
NSE solutions are much more economical, principally because of the much 
smaller number of free parameters which must be evolved, they are 
applicable to only a few situations.  The principle difficulty with 
evolution via rate equation is the computational cost, which results 
primarily from the stiffness of the system of nuclear reactions.  This 
extreme stiffness requires implicit solution and has thus far generally 
precluded the use of integration methods which do not rely on the Jacobian 
matrix.  Such non-Jacobian methods remain highly sought after as a means to 
reduce the computational cost of nucleosynthesis calculations.  For 
Jacobian based integration methods, there remain considerable economies to 
be gained by taking better advantage of the sparse nature of the Jacobian.

Physically motivated approaches can also be extremely useful in reducing the 
computational cost.  One such is the use of local equilibria to reduce the 
size of the system of rate equations (and reduce problems with matrix 
singularity). Methods based on local 
equilibria are applicable to many situations where global equilibrium has 
not been achieved.  Though the use of hybrid equilibrium networks is in its 
infancy, it seems that in most of the situations where one would have 
heretofore used a large network coupled with hydrodynamics, the hybrid 
equilibrium networks provide sufficient accuracy and considerable reduction 
in the computational cost.

\begin{ack}
As a review article, this article naturally owes much to the innumerable 
investigators who have devoted at least part of their life's work to our 
understanding of nucleosynthesis in astrophysical environments.  The 
authors would like to single out discussions with W.D. Arnett, G. Bazan, A.G.W.
Cameron, R.D. Hoffman, F. K\"appeler, B..S. Meyer, E. M\"uller, K. Nomoto, 
M. Wiescher and S.E. Woosley which were 
particularly useful in the preparation of this review.  We would also like 
to thank the Institute of Theoretical Physics at the Univ.  of California, 
Santa Barbara, for its hospitality and inspiration during the supernova 
program, support for which was provided under NSF grant No.  PHY94-07194.  
This work was supported by the U.S. Department of Energy under contract 
DE-FG02-96ER40983 (Joint Institute for Heavy Ion Research) and 
DE-AC05-96OR22464 (with Lockheed Martin Energy Research Corp) and by the 
Swiss Nationalfonds (grant 20-53798.98).
\end{ack}

%
~
%


\begin{thebibliography}{99}

\bibitem{BBFH57}                                                                
    Burbidge, E.M., Burbidge, G.R., Fowler, A.A.,  Hoyle, F. 1957,            
    \revmodphys, 29, 547                                                        
\bibitem{Came57}                                                                
    Cameron, A.G.W. 1957, Atomic Energy of Canada, Ltd., CRL-41                 
\bibitem{Came82}                                                                
    Cameron, A.G.W. 1982, in {\em Essays in Nuclear Astrophysics}, ed.          
    C.A. Barnes, D.D. Clayton,  D.N. Schramm (Cambridge Univ.:Cambridge), 23  
\bibitem{AnGr89}                                                                
    Anders, E., Grevesse, N. 1989, \geocosa, 53, 197                            
\bibitem{Clay83}                                                               
    Clayton, D.D. 1968, 1983, {\em Principles of Stellar Evolution and          
    Nucleosynthesis}, (Univ. of Chicago:Chicago)                                
\bibitem{KiWe90}                                                                
    Kippenhahn, R.,  Weigert, A. 1990 {\em Stellar Structure and              
    Evolution}, (Springer-Verlag:Berlin)                                        
\bibitem{Arne96}                                                               
    Arnett, W.D. 1996, {\em Supernovae and Nucleosynthesis}, (Princeton         
    Univ.:Princeton)                                                            
\bibitem{Baea82}                                                                
    Bahcall, J.N., H\"ubner, W.F., Lubow, S.H., Parker, P.D.,  Ulrich,        
    R.K. 1982, \revmodphys, 54, 767                                             
\bibitem{ABKW89}                                                               
    Arnett, W.D., Bahcall, J.N., Kirshner, R.P., Woosley, S.E. 1989, \araa,     
    27, 629                                                                     
\bibitem{WoAC73}                                                                
    Woosley, S.E., Arnett, W.D., Clayton, D.D. 1973, \apjs, 26, 231             
\bibitem{HiTh96}                                                               
     Hix, W.R.,  Thielemann F.-K. 1996, \apj, 460, 869                          
\bibitem{ArTh85}                                                               
    Arnett, W.D., Thielemann, F.-K. 1985, \apj, 295, 589                        
\bibitem{ThAr85}                                                               
    Thielemann, F.-K., Arnett, W.D. 1985, \apj, 295, 604                        
\bibitem{WoWe95}                                                               
    Woosley, S.E., Weaver, T.A. 1995, \apjs, 101, 181                           
\bibitem{Noea97}                                                               
    Nomoto, K., Hashimoto, M., Tsujimoto, T., Thielemann, F.-K., Kishimoto,     
    N., Kubo, Y., Nakasato, N. 1997, \nphysa, 161, 79c                          
\bibitem{KaBW89}                                                                
    K\"appeler, F., Beer, H., Wisshak, K. 1989, \repprogphys 52, 945            
\bibitem{Kaea94}                                                                
    K\"appeler, F. et al. 1994, \apj, 437, 396                                  
\bibitem{WVKK97}                                                               
    Wisshak, K., Voss, F., K\"appeler, F., Kerzakov, I. 1997, \nphysa, 621, 270c
\bibitem{GaBL97}                                                                
    Gallino, R., Busso, M., Lugarno, M. 1997, in {\em Astrophysical Implications
    of the Laboratory Study of Presolar MAterials}, eds. T. Bernatowicz, E. 
    Zinner, (AIP:New York), p. 115                              
\bibitem{Mezz99}                                                                
    Mezzacappa, A., \jcam, in press                                                 
\bibitem{Woos88}                                                               
    Woosley, S.E. 1988, \apj, 330, 218                                          
\bibitem{Mccr93}                                                               
    McCray, R. 1993, \araa, 31, 175                                             
\bibitem{Trur85}                                                                
    Truran, J.W. 1985, \arnp, 34, 53                                            
\bibitem{Trim91}                                                               
    Trimble, V. 1991, \aapr, 3, 1                                               
\bibitem{Arne95}                                                               
    Arnett, W.D. 1995, \araa, 33, 115                                           
\bibitem{ThHN90}                                                               
    Thielemann, F.-K., Hashimoto, M., Nomoto, K. 1990,  \apj, 349, 222          
\bibitem{ThNH96}                                                                
    Thielemann, F.-K., Nomoto, K.,   Hashimoto, M. 1996, \apj, 460, 408       
\bibitem{Came89}                                                               
    Cameron, A.G.W. 1989, in {\em Cosmic Abundances of Matter}, ed.             
    C.J. Waddington, AIP Conf. Proc. 183, 349                                   
\bibitem{Meye89}                                                                
    Meyer, B.S. 1989, \apj, 343, 254                                            
\bibitem{LMRS77}                                                                
    Lattimer, J.M., Mackie, F., Ravenhall, D.G., Schramm, D.N. 1977, \apj,      
    213, 225                                                                    
\bibitem{ELPS89}                                                                
    Eichler, D., Livio, M., Piran, T., Schramm, D.N. 1989, \nat, 340, 126       
\bibitem{RLTD98}                                                                
    Rosswog, S., Liebend\"orfer, M., Thielemann, F.-K., Davies, M.B., Benz, W.,
    Piran, T. 1998, \aap, 341 499                     
\bibitem{WoHo92}                                                                
    Woosley, S.E., Hoffman, R.D. 1992, \apj, 395, 202                           
\bibitem{Hoea97}                                                                
    Hoffman, R.D., Woosley, S.E., Qian, Y.-Z. 1997, \apj, 482, 951              
\bibitem{KBTM93}                                                                
    Kratz, K.-L., Bitouzet, J.-P., Thielemann, F.-K., M\"oller, P., Pfeiffer, B.
    1993, \apj, 402, 216                             
\bibitem{FKRT97}                                                                
    Freiburghaus, C., Kolbe, E., Rauscher, T., Thielemann, F.-K., Kratz,        
    K.-L., Pfeiffer, B. 1997, \nphysa, 621, 405c                                
\bibitem{Chan35}                                                                
    Chandrasekhar, S. 1935, \mnras, 95, 207                                     
\bibitem{NoTY84}                                                                
    Nomoto, K., Thielemann, F.-K.,  Yokoi, K. 1984, \apj, 286, 644            
\bibitem{NiWo97}                                                                
    Niemeyer, J.C.,  Woosley, S.E. 1997, \apj,  475, 740                      
\bibitem{KhOW97}                                                                
    Khokhlov, A.M., Oran, E.S.,  Wheeler, J.C. 1997, \apj, 478, 678           
\bibitem{HoWT98}                                                                
    H\"oflich, P., Wheeler, J.C., Thielemann, F.-K.  1998, \apj, 495, 617       
\bibitem{LiAr95}                                                                
    Livne, E.,  Arnett, W.D. 1995, \apj, 452, L62                             
\bibitem{Woos97a}                                                               
    Woosley, S.E. 1997, in {\em Thermonuclear Supernovae}, eds. P.              
    Ruiz-Lapuente, R. Canal, J. Isern, Kluwer Academic Publishers, p. 313       
\bibitem{SuNo80}                                                                
    Sugimoto, D.  Nomoto, K. 1980, \spscirev, 25, 155                         
\bibitem{Stea93}                                                                
    Starrfield, S., Truran, J.W., Politano, M.,  Sparks, W.M., Nofar, I.        
     Shaviv, G. 1993, \physrep, 227 155                                       
\bibitem{Coea95}                                                                
    Coc, A. Mochkovitchm R. Oberto, Y. Thibaud, J.-P.  Vangioni-Flam,         
    E. 1995, \aap, 299, 479                                                     
\bibitem{RFRT97}                                                                
    F. Rembges, C. Freiburghaus, T. Rauscher, F.-K. Thielemann, H. Schatz,      
    M. Wiescher 1997 , \apj, 484 412                                             
\bibitem{JoHe98}                                                                
    Jose, J.,  Hernanz, M. 1998, \apj, 494, 680                               
\bibitem{WaWo81}                                                                
    Wallace R. K.  Woosley, S. E. 1981, \apjs, 45, 389                        
\bibitem{TaWL96}                                                                
    Taam, R.E., Woosley, S.E.  Lamb, D.Q. 1996, \apj, 459, 271                
\bibitem{SAGW98}
    Schatz, H., Aprahamian, A,. G\"orres, J., Wiescher, M., Rauscher, T.,
    Rembges, J.F., Thielemann, F.-K., Kratz, K.-L., Pfeiffer, B.,
    M\"oller, P., Herndl, B., Brown, B.A. 1998, \physrep, 294, 167
\bibitem{Wago73}                                                                
    Wagoner, R.V. 1973, \apj, 179, 343                                          
\bibitem{ThSO94}                                                                
    Thomas, D., Schramm, D.N., Olive, K.A., 1994, \apj, 430, 291 
\bibitem{SmKM93}
    Smith, M.S., Kawano, L.H.,  Malaney, R.A. 1993, \apjs, 85, 219
\bibitem{OlSS97}
     Olive, K.A., Steigman, G., Skillman, F.D. 1997, \apj, 483, 788         
\bibitem{RoRo88}                                                                
    Rolfs, C., Rodney, W.S. 1988, {\em Cauldrons in the Cosmos},                
    (Univ. of Chicago:Chicago)                                                  
\bibitem{Woos86}
    Woosley, S.E. 1986, in {\em Nucleosynthesis and Chemical Evolution},
    16th Advanced Course of the Swiss Society of Astrophysics and
    Astronomy, ed. B. Hauck, A. Maeder,  G. Meynet (Geneva:Geneva Obs.), 1
\bibitem{ThNH94}                                                                
    Thielemann, F.-K., Nomoto, K., Hashimoto, M. 1994, in {\em Supernovae,      
    Les Houches, Session LIV}, eds. S. Bludman, R. Mochkovitch, J.              
    Zinn-Justin, Elsevier, Amsterdam, p. 629                                    
\bibitem{FoCZ67}                                                                
    Fowler, W.A., Caughlan, G.E., Zimmermann, B.A. 1967, \araa, 5, 525          
\bibitem{Waea97}                                                                
    Wallerstein, G. et al. 1997, \revmodphys, 69, 995                           
\bibitem{KaTW98}
    K\"appeler, F., Thielemann, F.-K.  Wiescher, M. 1998, \arnp, 48, 175
\bibitem{CaFo88}                                                                
    Caughlan, G.R., Fowler, W.A. 1988, \andt, 40, 283                           
\bibitem{ArGJ99}                                                                
    Arnould, M., Goriely, S., \& Jorissen, A. 1999, \aap, in press    
\bibitem{BaKa87}                                                                
    Bao, Z.Y., K\"appeler, F. 1987, \andt, 36, 411                              
\bibitem{BeVW92}                                                               
    Beer, H., Voss, F., Winters, R.R. 1992, \apjs, 80, 403                      
\bibitem{WiGT88}                                                               
    Wiescher, M., G\"orres, J., Thielemann, F.-K. 1988, \apj, 326, 384          
\bibitem{WGGB89}                                                               
    Wiescher, M., G\"orres, J., Graaf, S, Buchmann, L., Thielemann, F.-K.       
    1989, \apj, 343, 352                                                        
\bibitem{TSOF93}                                                                
    Thomas, D., Schramm, D.N., Olive, K.A., Fields, B.D. 1993, \apj, 406, 569   
\bibitem{Raea94}                                                                
    Rauscher, T., Applegate, J.H., Cowan, J.J., Thielemann, F.-K., Wiescher,    
    M. 1994, \apj, 429, 499                                                     
\bibitem{HWFZ76}                                                                
    Holmes, J.A., Woosley, S.E., Fowler, W.A., Zimmerman, B.A. 1976, \andt,     
    18, 306                                                                     
\bibitem{WFHZ78}                                                                
    Woosley, S.E., Fowler, W.A., Holmes, J.A., Zimmerman, B.A. 1978, \andt,     
    22, 371                                                                     
\bibitem{ThAT87}                                                                
    Thielemann, F.-K., Arnould, M., Truran, J.W. 1987, in {\em Advances in      
    Nuclear Astrophysics}, ed. E. Vangioni-Flam, Gif sur Yvette, Editions       
    Fronti\`ere, p.525                                                          
\bibitem{CoTT91}                                                                
    Cowan, J.J., Thielemann, F.-K., Truran, J. W. 1991, \physrep, 208, 267      
\bibitem{RaTK97}                                                                
    Rauscher, T., Thielemann, F.-K., Kratz, K.-L. 1997, \prc, 56, 1613          
\bibitem{RaTh98}                                                                
    Rauscher, T.,  Thielemann, 1998, in {\em Second Oak Ridge                 
    Symposium on Atomic and Nuclear Astrophysics}, ed. A. Mezzacappa,           
    (Bristol:IoP) 519
\bibitem{SFKR98}
    Somorjai, E., \etal 1998, \aap, 333, 1112                                                      
\bibitem{Salp54}                                                                
    Salpeter, E.E. 1954 \ausjphys, 7, 373                                       
\bibitem{SavH69}                                                               
   Salpeter, E.E., van Horn, H.M. 1969, \apj, 155, 183                          
\bibitem{ThTr87}                                                               
    Thielemann, F.-K., Truran, J.W. 1987, in {\em Advances in Nuclear           
    Astrophysics}, eds. E. Vangioni-Flam et al., Editions Fronti\`eres,         
    Gif sur Yvette, p. 541                                                                       
\bibitem{Ichi93}                                                               
    Ichimaru, S. 1993, \revmodphys, 65, 255                                                             
\bibitem{BrSa97}                                                               
    Brown, L.S., Sawyer, R.F. 1997, \revmodphys, 69, 411                        
\bibitem{Ichi96}                                                               
        Ichimaru, S. 1996, \pubastjap, 48, 613                                  
\bibitem{FuFN80}                                                               
    Fuller, G.M., Fowler, W.A., Newman, M. 1980, \apjs, 42, 447                 
\bibitem{FuFN82}                                                               
    Fuller, G.M., Fowler, W.A., Newman, M. 1982, \apjs, 48, 279                 
\bibitem{FuFN85}                                                               
    Fuller, G.M., Fowler, W.A., Newman, M. 1985, \apj, 293, 1                   
\bibitem{Taea89}                                                                
    Takahara, M., Nino, M., Oda, T., Muto, K., Wolters, A.A., Claudemans,       
    P.W.M., Sato, K. 1989, \nphysa, 504, 167                                    
\bibitem{Deea98}                                                                
    Dean, D.J., Langanke, K., Chatterjee, L., Radha, P.B., Strayer, M.R.        
    1998, \prc, 58, 536                                                    
\bibitem{AFWH94}                                                               
    Aufderheide, M., Fushiki, I., Woosley, S.E., Hartmann, D. 1994,             
    \apjs, 91, 389                                                              
\bibitem{SuSR97}                                                                
    Sutaria, F.K., Sheikh, J.H., Ray, A. 1997, \nphysa, 621, 375c               
\bibitem{TaYK73}                                                                
    Takahashi, K., Yamada, M., Kondo, Z. 1973, \andt, 12, 101                   
\bibitem{KlMO84}                                                               
    Klapdor, H.V., Metzinger, J., Oda, T. 1984, \andt, 31, 81                   
\bibitem{TaYo88}                                                                
    Takahashi, K., Yokoi, K. 1988, \andt, 36, 375                               
\bibitem{SBMK89}                                                               
    Staudt, A., Bender, E., Muto, K., Klapdor, H.V. 1989, \zphys, A334, 47      
\bibitem{SBMK90}                                                               
    Staudt, A., Bender, E., Muto, K., Klapdor, H.V. 1990, \andt, 44, 79         
\bibitem{MoRa90}                                                                
    M\"oller, P., Randrup, J. 1990, \nphysa, 514, 1                             
\bibitem{HSMK92}                                                               
    Hirsch, M., Staudt, A., Muto, K., Klapdor-Kleingrothaus, H.V. 1992,         
     \nphysa, 535, 62                                                           
\bibitem{PfKr96}                                                                
    Pfeiffer, B., Kratz, K.-L. 1996, KCh Mainz Report, unpublished              
\bibitem{MoNK97}                                                                
    M\"oller, P., Nix, J.R., Kratz, K.-L. 1997, \andt, 66, 131                  
\bibitem{Borz97}                                                                
    Borzov, I.N. 1996, \zphys, A335, 125; 1997, \nphysa, 621, 307c              
\bibitem{OdHM94}
    Oda, T., Hino, M., Minto, K. 1994, \andt, 56, 231
\bibitem{FuMe95}                                                                
    Fuller, G., Meyer, B.S. 1995, \apj, 453, 792                                
\bibitem{WHHH90}                                                               
    Woosley, S.E., Hartmann, D., Hoffman, R.B., Haxton, W.C. 1990, \apj,        
    356, 272                                                                    
\bibitem{KKLT92}                                                                
    Kolbe, E., Krewald, S., Langanke, K., Thielemann, F.-K. 1992, \nphysa,      
    540, 599                                                                    
\bibitem{KLKT93}                                                                
    Kolbe, E., Langanke, K., Krewald, S., Thielemann, F.-K. 1993, \physrep,     
    227, 37                                                                     
\bibitem{KLTV95}                                                                
    Kolbe, E., Langanke, K., Thielemann, F.-K., Vogel, P. 1995, \prc, 52, 3437  
\bibitem{Qiea96}                                                                
    Qian, Y.-Z., Woosley, S.E., Haxton, W.C., Langanke, K., Vogel, P. 1996,     
    \prc, 55, 1532                                                              
\bibitem{NoTM85}                                                               
    Nomoto, K., Thielemann, F.-K., Miyaji, S. 1985, \aap, 149, 239              
\bibitem{GoWT95}                                                               
    G\"orres, J., Wiescher, M., Thielemann, F.-K. 1995, \prc, 51, 392           
\bibitem{Muel98}
    M\"uller, E. 1998, in {\em Computational Methods for Astrophysical Fluid
    Flow}, eds. O. Steiner  A. Gautschy, (Springer-Verlag:Berlin), 343 
\bibitem{AuWa95}                                                                
    Audi, G., Wapstra, A.H. 1995, \nphysa, 595, 409                             
\bibitem{Muel86}
    M\"uller, E. 1986, \aap, 162, 103
\bibitem{ArTr69}
    Arnett, W.D., Truran, J.W. 1969, \apj, 157, 339; 1369
\bibitem{WeZW78}
    Weaver, T.A., Zimmerman, G.B., Woosley, S.E. 1978, 225, 1021
\bibitem{BeHT89}
    Benz, W., Hilles, J.G., Thielemann, F.-K. 1989, \apj, 342 986
\bibitem{FrMA89}
    Fryxell, B., M\"uller, E.,  Arnett, W.D. 1989, {\em Max-Planck-Institut
    f\"ur Astophysik Preprint}, 449, Garching
\bibitem{Larr91}
    Larrouturou, B. 1991, \jcompphys, 95, 59
\bibitem{PlMu99}
    Plewa, T.,  M\"uller, E. 1999, \aap, 342, 179
\bibitem{BaAr98}
     Bazan, G., Arnett, W.D. 1998, \apj 496 316       
\bibitem{Gear71}                                                                
    Gear, C.W. 1971, {\em Numerical Initial Value Problems in Ordinary          
    Differential Equations}, (Prentice-Hall:Englewood Cliffs, NJ)               
\bibitem{Lamb80}                                                                
    Lambert, J.D. 1980, in {\em Computational Techniques for Ordinary           
    Differential Equations}, eds. I., Gladwell, D.K. Sayars,                  
    (Academic:NY), p.19                                                         
\bibitem{OrBo87}                                                                
    Oran, E.S., Boris, J.P. 1987, {\em Numerical Simulation of                
    Reactive Flow}, (Elsevier:NY)                                               
\bibitem{NumRec}                                                                
    Press, W.H., Teukolsky, S.A., Vetterling, W.T., Flannery, B.P.  1992,     
    Numerical Recipes (Cambridge:Cambridge Univ.), second edition               
\bibitem{Timm99}
    Timmes, F.X. 1999, \apjs, in press
\bibitem{CoCT83}
    Cowan, J.J., Cameron, A. G.W., Truran, J.W. 1983, \apj, 265, 429
\bibitem{FRRK99}
    Freiburghaus, C., Rembges, F., Rauscher, T., Kolbe, E., Thielemann, F.-K.,
    Kratz, K.-L., Pfeiffer, B., Cowan, J.J. 1999, \apj, 516, 381
\bibitem{ClTa65}                                                                
    Clifford, F.E., Tayler, R.J. 1965, \memras, 69, 21                       
\bibitem{Came79}                                                                
    Cameron, A.G.W. 1979, \apjl, 230, L53                                       
\bibitem{HaWE85}                                                                
    Hartmann, D., Woosley, S.E., El Eid, M.F.  1985, \apj, 297, 837          
\bibitem{HTFT99}                                                                
    Hix, W.R., Thielemann F.-K., Fushiki, I., Truran, J.W. 1999, in
    preparation                                                                   
\bibitem{LaLi58}                                                                
    Landau, L.D., Lifshitz, E.M. 1958, {\em Theoretical Physics V            
    (Statistical Mechanics)}, (Pergamon:London)                                 
\bibitem{MeKC98}                                                                
    Meyer, B.S., Krishnan, T.D., Clayton, D.D. 1998, \apj, 498, 808         
\bibitem{HiTh99}                                                               
     Hix, W.R.,  Thielemann F.-K. 1999, \apj, 511, 862                          
\bibitem{Krea93}                                                               
    Kratz, K.-L., Bitouzet, J.-P., Thielemann, F.-K., M\"oller, P.,             
    Pfeiffer, B. 1993,  \apj, 402, 216                                          
\bibitem{Boea96}                                                                
    Bouquelle, V., Cerf, N., Arnould, M., Tachibana, M., Goriely, S. 1996,  
    \aap, 305, 1005                                                             
\bibitem{BoCF68}                                                                
    Bodansky, D., Clayton, D.D., Fowler, W.A.  1968, \apjs, 16, 299          
\bibitem{HiFT99}                                                                
    Hix, W.R., Freiburghaus, C.,  Thielemann F.-K. 1999, \apj submitted       
\bibitem{HKWT98}                                                                
    Hix, W.R., Khokhlov, A.M., Wheeler, J.C., Thielemann F.-K. 1998, \apj,    
    503, 332                                                                   
\bibitem{FoHo64}                                                                
    Fowler, W.A., Hoyle, F. 1964, \apjs, 9, 201                              

\end{thebibliography}
\end{document}